\documentclass{aa}
\newcommand{\xmm}{\textit{XMM-Newton}\xspace}
\newcommand{\chandra}{\textit{Chandra}\xspace}
\newcommand{\rosat}{\textit{ROSAT}\xspace}
\newcommand{\suzaku}{\textit{Suzaku}\xspace}
\usepackage{natbib,twoopt}
\usepackage{xcolor}
\usepackage[breaklinks=true, colorlinks=true, linkcolor=blue, citecolor=blue]{hyperref} 
\bibpunct{(}{)}{;}{a}{}{,}             
\makeatletter
\newcommandtwoopt{\citeads}[3][][]{\href{http://adsabs.harvard.edu/abs/#3}%
    {\def\hyper@linkstart##1##2{}%
     \let\hyper@linkend\@empty\citealp[#1][#2]{#3}}}
  \newcommandtwoopt{\citepads}[3][][]{\href{http://adsabs.harvard.edu/abs/#3}%
    {\def\hyper@linkstart##1##2{}%
     \let\hyper@linkend\@empty\citep[#1][#2]{#3}}}
  \newcommandtwoopt{\citetads}[3][][]{\href{http://adsabs.harvard.edu/abs/#3}%
    {\def\hyper@linkstart##1##2{}%
     \let\hyper@linkend\@empty\citet[#1][#2]{#3}}}
  \newcommandtwoopt{\citeyearads}[3][][]%
    {\href{http://adsabs.harvard.edu/abs/#3}
    {\def\hyper@linkstart##1##2{}%
     \let\hyper@linkend\@empty\citeyear[#1][#2]{#3}}}
\makeatother
\usepackage{graphicx}
\usepackage{multirow}
\usepackage{comment}
\usepackage{siunitx}  
\usepackage{txfonts}
\usepackage{subfig}
\usepackage{sidecap}
\sidecaptionvpos{figure}{c}
\begin{document}

   \title{The SRG/eROSITA All-Sky Survey:\\Large-scale view of the Centaurus cluster}
   \subtitle{}
   \titlerunning{eROSITA large-scale view of the Centaurus cluster}

   \author{Angie Veronica
          \inst{1}
          \and
          Thomas H. Reiprich\inst{1}
          \and
          Florian Pacaud\inst{1}
          \and
          Jeremy S. Sanders\inst{2}
          \and
          Efrain Gattuzz\inst{2}
          \and
          Michael ~C.~H.~Yeung\inst{2}
          \and
          Esra Bulbul\inst{2}
          \and
          Vittorio Ghirardini\inst{4,2}
          \and
          Ang Liu\inst{2}
          \and
          Caroline Mannes\inst{1}
          \and
          Alexander Morelli\inst{1}
          \and
          Naomi Ota\inst{1,3}
          }

   \institute{Argelander-Institut f\"ur Astronomie (AIfA), Universit\"at Bonn, Auf dem H\"ugel 71, 53121 Bonn, Germany\\
        \email{averonica@astro.uni-bonn.de}
        \and
        Max-Planck-Institut f\"ur extraterrestrische Physik, Gießenbachstraße 1, 85748 Garching, Germany
        \and
        Nara Women's University, Kitauoyanishi-machi, Nara, 630-8506, Japan
        \and
        INAF, Osservatorio di Astrofisica e Scienza dello Spazio, via Piero Gobetti 93/3, 40129 Bologna, Italy
        }
   \date{Accepted: December 17, 2024}

 
  \abstract
   {The Centaurus cluster is one of the brightest and closest clusters. Previous comprehensive studies were done only in the brightest part ($r<30'$) where the centers of the main substructures (Cen 30 and Cen 45) are located. Only a small fraction of the outskirts has been studied.}
   {This work aims to characterize the ICM morphology and properties of the Centaurus cluster out to the radius within which the density is 200 times the critical density of the Universe at the redshift of the cluster, $R_{200}~(91')$.}
   {We utilized the combined five SRG/eROSITA All-Sky Survey data (eRASS:5) to perform X-ray imaging and spectral analyses in various directions out to large radii. We employed some image manipulation methods to enhance small and large-scale features. Surface brightness profiles out to $2R_{200}$ were constructed to quantify the features. We acquired gas temperature, metallicity, and normalization per area profiles out to $R_{200}$. We compared our results with previous Centaurus studies, cluster outskirts measurements, and simulations. Comprehensive sky background analysis was done across the FoV, in particular, to assess the variation of the eROSITA Bubble emission that partially contaminates the field.}
   {The processed X-ray images show the known sloshing-induced structures in the core, such as cool plume, cold fronts, and ram-pressure-stripped gas. The spectra in the core ($r\leq11~\mathrm{kpc}$) are better described with a two-temperature model than an isothermal model. With this 2T analysis, we measured lower temperature from the cooler component ($\sim\!1.0~\mathrm{keV}$) and higher metallicity ($\sim\!1.6Z_\odot$), signifying an iron bias. In the intermediate radial range, the temperature peaks at $\sim\!3.6~\mathrm{keV}$, and we observed prominent surface brightness and normalization per area excesses in the eastern sector (Cen 45 location). Temperature enhancements near the location of Cen 45 imply that the gas is shock-heated due to the interaction with the Cen 30. We reveal that the eastern excess emission extends even further out, reaching $R_{500}$. The peak excess of normalization is located at $\sim\!23'$ from the center ($8'$ behind the center of Cen 45) with a $45\%$ and $7.7\sigma$ above the full azimuthal value. This might be the tail/ram-pressure-stripped gas from Cen 45. There is a temperature decrease of a factor about $2-3$ from the peak to the outermost bin at $R_{500}-R_{200}$. We found good agreement between the outer temperatures ($r>R_{2500}$) with the temperature profile from simulations and temperature fit from \suzaku cluster outskirts measurements.
   We detect significant surface brightness emission to the sky background level out to $R_{200}$ with a $3.5\sigma$ and followed with $2.9\sigma$ at $1.1R_{200}$. The metallicity at $R_{500}-R_{200}$ is low but within the ranges of other outskirts studies.}
   {We present the first measurement of ICM morphology and properties of Centaurus cluster sampling the whole azimuth beyond $\sim\!30'$, increasing the probed volume by a factor of almost 30. While the cluster core is rich in features as the result of AGN feedback and sloshing, the cluster outskirts temperature of Centaurus follows the temperature profile of clusters in simulations, as well as temperature fit from other cluster outskirts measurements.}

   \keywords{Galaxies: clusters: individual: Abell 3526, Centaurus -- X-rays: galaxies: clusters -- Galaxies: clusters: intracluster medium}

   \maketitle

\section{Introduction}
\begin{SCfigure*}[][!h]
\includegraphics[width=0.75\textwidth]{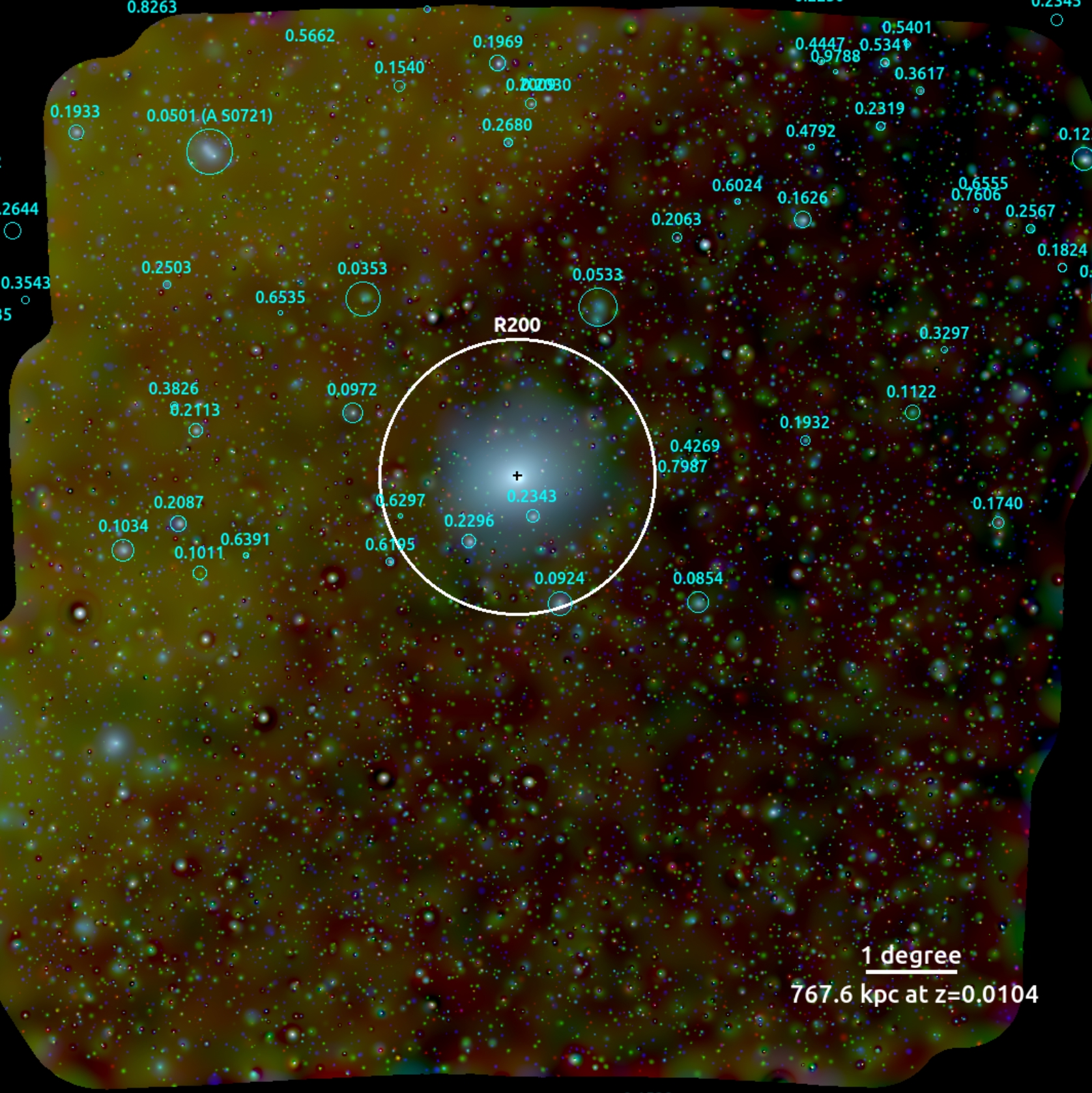}
\hspace{5pt}
\caption{Composite image of the large scale view of Centaurus (A3526) using eRASS:5 data. Red, green, and blue channels are in the $0.2-0.6~\mathrm{keV}$, $0.6-1.0~\mathrm{keV}$, and $1.0-2.3~\mathrm{keV}$ energy bands, respectively. The image in each channel is PIB-subtracted, exposure-corrected, Galactic-absorption corrected, and wavelet-filtered. The cyan circles are the clusters from the eRASS1 cluster catalog \citep{Bulbul_2024, Kluge_2024}.}
\label{fig:rgb}
\end{SCfigure*}

The Centaurus cluster (Abell 3526) is nearby ($z=0.0104$) and one of the X-ray brightest clusters with a $0.1-2.4~\mathrm{keV}$ luminosity within the $R_{500}$ of $6.9\times10^{43}~\mathrm{erg~s}^{-1}$ \citep{Piffaretti_2011}. It has been extensively studied by X-ray instruments, for example, \textit{ASCA} \citep[e.g.,][]{Churazov_1999, Dupke_2001, Furusho_2001}, \rosat \citep[e.g.,][]{Allen_1994, Churazov_1999, Ikebe_1999}, \suzaku \citep[e.g.,][]{Ota_2007, Walker_2013Suzaku, OtaYoshida_2016}, \xmm \citep[e.g.,][]{Walker_2013, Fukushima_2022, Gatuzz_2022, Gatuzz_2023}, and \chandra \citep[e.g.,][]{SandersFabian_2002, Sanders_2016, Lakhchaura_2019}. The X-ray emission of the Centaurus cluster appears smooth with an elliptical morphology \citep{Allen_1994} and it is peaked at its brightest cluster galaxy (BCG), NGC 4696 \citep{Dupke_2001}.
\par
However, early studies of the line-of-sight velocity distribution of the member galaxies show a bimodality, indicating the presence of two substructures with a velocity difference of $1500~\mathrm{km~s}^{-1}$ \citep{Lucey_1986, LuceyIII_1986}. The main substructure, which sits in the center of the cluster, is called Cen 30 and is centered on the NGC 4696 galaxy. The other substructure, located $\sim\!15'$ east of the center, is called Cen 45 and is centered on the NGC 4709 galaxy. Based on the results of $U-V$ color-magnitude relations, luminosity functions, and galactic radius distributions of both substructures, \cite{Lucey_1986} concluded that Cen 30 and Cen 45 belong to the same system, which also indicates that the Cen 45 is being accreted toward Cen 30. The interaction between the two substructures is further supported by findings of temperature excess near the Cen 45 location, which can be explained by shock heating \citep{Churazov_1999, Walker_2013}. 
\par
The central region of the cluster is filled with low-temperature and metal-rich gas \citep[e.g.,][]{Ikebe_1999, SandersFabian_2002} and numerous structures (soft filaments, cold fronts, cavities, etc.) that are related to the AGN feedback of the BCG and sloshing motions \citep[e.g.,][]{Taylor_2006, Sanders_2016}.
\par
Being nearby, the Centaurus cluster has a relatively large apparent size in the sky ($R_{500}\approx59'$). Hence, most of the X-ray studies are directed to probe the inner $r<30'$ region due to the field-of-view (FoV) limitation. An attempt to probe the outskirts of the Centaurus cluster was done in \cite{Walker_2013Suzaku}, with a stripe of six \suzaku observations along the north-west direction (avoiding the known cold fronts and bright point sources) up to $0.95R_{200}$. In this work, we present a large-scale view of the Centaurus cluster using the eROSITA All-Sky Survey data \citep[eRASS;][]{Predehl_2021, Merloni_2024, Bulbul_2024}. We present the first ICM profiles (surface brightness, temperature, metallicity, and normalization per area) out to $R_{200}$ sampling the full azimuth, as well as four different sectors divided based on the ellipticity of the cluster (semi-major/merger axis versus semi-minor axis). Therefore, we increase the probed volume of the Centaurus cluster by a factor of almost $30$.
\par
The structure of this paper is the following: in Sect.~\ref{sec:drsteps}, the data reduction steps, as well as the analysis strategy for imaging and spectral analyses are described. In Sect.~\ref{sec:results}, the results are presented and discussed in Sect.~\ref{sec:discussion}. We close with the summary and conclusions in Sect.~\ref{sec:conclude}.
\par
Unless stated otherwise, all uncertainties are at the 68.3\% confidence interval. The assumed cosmology in this work is a flat $\Lambda$CDM cosmology, where the Hubble constant is $H_0=70~\mathrm{km~s^{-1}~Mpc^{-1}}$. At the assumed redshift of the Centaurus cluster, $z=0.0104$, $1''$ corresponds to 0.213 kpc.

\section{Data reduction and analysis}\label{sec:drsteps}
\begin{table}
\centering
\caption{Information on the Centaurus cluster (Abell 3526) used in the present work.}
\resizebox{\columnwidth}{!}
{\begin{tabular}{c c}
\hline
\hline
R.A., Dec. [J2000] & $192.200^{\circ}, -41.308^{\circ~a}$ \\
Redshift $z$ & $0.0104^b$\\
$R_{500}$ & 0.826 Mpc ($3538.50''$)$^a$ \\
$R_{200}$ & 1.271 Mpc ($5443.85''$)$^c$\\
\hline
\multicolumn{2}{l}{\footnotesize $^a$\cite{Piffaretti_2011}, $^b$\cite{Lucey_1986},}\\
\multicolumn{2}{l}{\footnotesize $^cR_{200}\approx R_{500}/0.65\approx R_{2500}/0.28$ \citep{Reiprich_2013}}\\
\hline
\hline
\end{tabular}}
\label{tab:Ceninfo}
\end{table}

We employed nine sky tiles of the combined first five eROSITA All-Sky Survey data (eRASS:5) to cover the Centaurus cluster out to $3\times R_{200}$. The data was processed using the c020 configuration. The sky tiles are 188129, 188132, 188135, 191129, 192132, 192135, 195129, 196132, and 197135. The Centaurus cluster is mainly located at sky tile 192132. The data reduction steps were performed utilizing the extended Science Analysis Software (\texttt{eSASS}, \citealt{Brunner_2022}) version \texttt{eSASSusers\_211214} (eSASS4DR1\footnote{\href{https://erosita.mpe.mpg.de/dr1/eSASS4DR1/}{https://erosita.mpe.mpg.de/dr1/eSASS4DR1/}}). The data reduction started with the removal of unwanted pixels, such as bad pixels and strongly vignetted corners of the FoV by specifying \texttt{flag=0xe00fff30} in the \texttt{eSASS} task \texttt{evtool}. Afterwards, a light curve was generated for each sky tile by the \texttt{flaregti} task. We specified the minimum energy limit for the lightcurve creation of 5 keV (\texttt{pimin=5000}). A $3\sigma$ threshold was calculated from the lightcurve and supplied to the second run of \texttt{flaregti} by the parameter \texttt{threshold}. Two additional parameters were supplied, namely, the diameter of source extraction area in arcseconds unit (\texttt{source\_size=150}) and the number of grid points per dimension (\texttt{gridsize=26}). A \texttt{FLAREGTI} extension was generated for each sky tile and applied to the event file through \texttt{evtool} task by adding the parameter \texttt{gti="FLAREGTI"}. In total, $2.1\%$ of the exposure was removed, implying that the sky tiles used were not affected by any flares in all five eRASS passes. The average effective exposure time in the field is around $750~\mathrm{s}$. We then combined the cleaned and filtered sky tiles and centered them at the cluster center (Table~\ref{tab:Ceninfo}). For the image correction, which includes particle-induced background (PIB) subtraction, exposure correction (including vignetting), and Galactic absorption correction across the FoV, we follow the steps described in \cite{Reiprich_2021} (see their Sect.~2.1 and 3.3) and \cite{Veronica_2024}. We summarize the steps below and refer to \cite{Reiprich_2021} for more details.
\par
For the imaging analysis, the energy bands of $0.2-2.3~\mathrm{keV}$ and $1.0-2.3~\mathrm{keV}$ (see Sect.~\ref{sec:erobubble}) were used. The eROSITA Telescope Modules (TMs) consist of five TMs with on-chip filter (TM1, 2, 3, 4, 6; the combination is referred to as TM8) and two without on-chip filter (TM5 and 7; the combination is referred to as TM9). The latter suffer from the optical light leak at the bottom part of their respective CCDs \citep{Predehl_2021}. Due to this, different lower energy limits were used for the two types of eROSITA TMs, namely 0.2 keV and 0.8 keV for TM8 and TM9, respectively. It was discovered that the Al-K$\alpha$ at $\sim\!$1.4 keV is higher in the filter-wheel closed (FWC) spectra than in the observations. In order to not overestimate the PIB modelled for the observation, we decided to exclude the $1.35-1.6~\mathrm{keV}$ band from the imaging analysis. The correction takes into account the different lower energy limits and the count rates of the final, all-sky tiles combined, and a fully corrected image corresponds to an effective area given by one TM with an on-chip filter in the energy band $0.2-2.3~\mathrm{keV}$ excluding the Al-K$\alpha$ line. For simplicity, we will refer to this energy band as the $0.2-2.3~\mathrm{keV}$ band.
\par
A relative Galactic absorption correction in the FoV was done utilizing the $N_\mathrm{H,tot}$ map of the Centaurus field obtained by adding the HI4PI all-sky Galactic neutral atomic hydrogen map \citep{HI4PI_2016} with $N_{\mathrm{H}_2}$ map calculated using the method described in \cite{Willingale_2013} through the Swift Galactic column density of Hydrogen tool\footnote{\href{https://www.swift.ac.uk/analysis/nhtot/}{https://www.swift.ac.uk/analysis/nhtot/}}. The eROSITA count rates of different column density values, $N_\mathrm{H}$, across the FoV were estimated by simulating a typical X-ray fore- and background model that consists of the unabsorbed Local Hot Bubble (LHB), the absorbed Milky Way Halo (MWH), and unresolved sources. For each pixel, the $N_\mathrm{H}$ correction value was calculated, namely the ratio of the expected eROSITA count rate at the given pixel to the count rate of the median $N_\mathrm{H}$ across the FoV. The absorption correction factor map was generated for each type of TMs and multiplied by its corresponding exposure map.
\par
To detect and generate the point source catalog, we followed the procedure described in \cite{Pacaud_2006} and \cite{Ramos-Ceja_2019}. The source detection was performed on a wavelet-filtered image by the Source Extractor software \citep[\texttt{SExtractor},][]{SExtractor}. The method detects not only point sources, but also extended sources (including small extended background sources) in the FoV. We used the obtained catalog to remove point sources and unwanted or unrelated sources in our imaging and spectral analyses. In Fig.~\ref{fig:rgb} and \ref{fig:conf}, we overplotted background clusters from the eRASS1 groups and clusters catalog \citep{Bulbul_2024}, which are included in our source catalog and thus removed from further analyses. In order to emphasize low surface brightness and large-scale features, we applied adaptive smoothing to the images. We utilized the Science Analysis Software (SAS; version 20.0.0) task \texttt{asmooth} with the parameter \texttt{smoothstyle=‘adaptive’}. A cheese mask was also supplied which allows us to mask out unwanted sources and refill the holes with the surrounding background values. Any adaptively-smoothed images are for visualization purposes only.

\subsection{Imaging analysis}\label{sec:imag_analysis}
\begin{figure}[!h]
\includegraphics[width=\columnwidth]{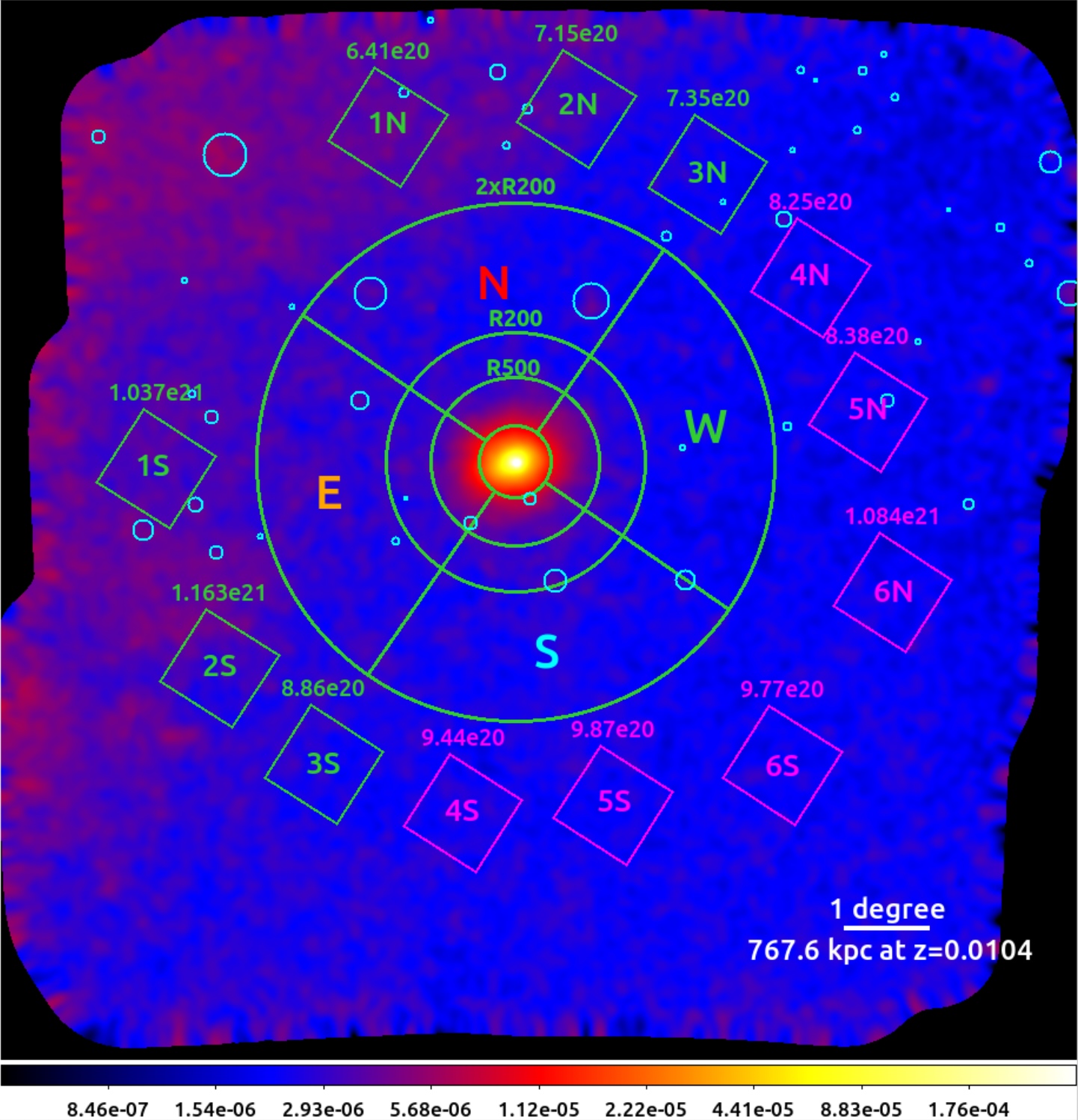}
\caption{eRASS:5 fully-corrected image in the $0.2-2.3~\mathrm{keV}$ band. The image has been adaptively smoothed with S/N set to 45 (see Sect.~\ref{sec:imag_analysis}). The green concentric circles are the characteristic radii of the Centaurus cluster and the four sectors used in the analyses are labelled. The boxes outside $2R_{200}$ are used for the CXB analysis described in Sect.~\ref{sec:erobubble}, as a result we decided to use the magenta boxes for CXB region for our science analyses. The values above the boxes are their median $N_\mathrm{H}$ values in atoms cm$^{-2}$. The cyan circles are the eRASS1 background clusters as shown in Fig.~\ref{fig:rgb}. In this image the emission from these clusters, as well as other unrelated sources have been removed and refilled.}
\label{fig:conf}
\end{figure}

We calculated surface brightness profiles of the Centaurus cluster out to $2\times R_{200}$ using eRASS:5 products (cleaned- and flare-filtered photon map, PIB map, fully-corrected exposure map) in the $0.2-2.3~\mathrm{keV}$ and $1.0-2.3~\mathrm{keV}$ energy bands (see Sect.~\ref{sec:erobubble}). To assess the morphology of the ICM, surface brightness profiles from four directions were constructed, namely north, west, south, and east. The opening angles of each sector were based on the position angle of the elliptical detection by \texttt{SExtractor}, such that the center of the opening angle of each sector is either the minor or major axis of the ellipse. We show the configurations of the opening angles of all the sectors, the characteristic radii\footnote{We convert the characteristic radii using the relations described in \cite{Reiprich_2013}, e.g., $R_{500}\approx0.65R_{200}$, $R_{100}\approx1.36R_{200}$, and $R_{2500}\approx0.28R_{200}$.}, and the cosmic X-ray background (CXB) regions used for our analyses in Fig.~\ref{fig:conf}.
All surface brightness profiles were centered at the center coordinates (Table \ref{tab:Ceninfo}).

\subsection{Spectral analysis}\label{sec:spectro}
The source and background spectra, as well as their ancillary response files (ARF) and response matrix files (RMFs), were extracted using \texttt{eSASS} \texttt{srctool} task. We performed all spectral fittings using the X-ray spectral fitting package \citep[\texttt{XSPEC},][]{XSPEC} version: 12.12.0. The spectral fitting procedure follows closely the steps described in \cite{Veronica_2024}, except for the additional CXB component to account for the eROSITA Bubble
emission (Sect.~\ref{sec:erobubble}) and the used PIB model version. We summarize the spectral fitting procedure below.
\par
The full spectral fitting model is
\begin{equation}
\begin{split}
\mathtt{Model =}&\quad\mathtt{(apec_1 + TBabs\times(apec_2 + apec_3 + powerlaw))}\\
&\quad\mathtt{+~TBabs\times apec_4 + PIB},\\
\end{split}
\label{eq:spectral_model}
\end{equation}
where the first term denotes the CXB components, including the unabsorbed Local Hot Bubble (LHB; \texttt{apec$\mathtt{_1}$}), and the absorbed Milky Way halo (MWH; \texttt{apec$\mathtt{_2}$}), the eROSITA Bubbles (\texttt{apec$\mathtt{_3}$}), and unresolved sources (PL; \texttt{powerlaw}). The \texttt{apec} component is described by five parameters: the plasma temperature ($k_\mathrm{B}T$), metal abundances ($Z$), redshift ($z$), and normalization ($norm$). The absorption along the line of sight by the Galactic column density is represented by \texttt{TBabs} \citep{Wilms_2000}. The $N_\mathrm{H}$ values used for the different regions are taken from the $N_\mathrm{H,tot}$ map (see Sect.~\ref{sec:drsteps}). The \texttt{powerlaw} component is described by photon index ($\Gamma$) and normalization parameters. For the spectral analysis, we always fixed $\Gamma=1.46$ \citep{Luo_2017}. The absorbed thermal emission from the cluster is described by the second term, \texttt{TBabs}$\times$\texttt{apec$\mathtt{_4}$}. The PIB is represented by the third term, \texttt{PIB}. The modeling of the PIB in the present work used the results of the analysis of the FWC data processed in configuration c020\footnote{\href{https://erosita.mpe.mpg.de/dr1/AllSkySurveyData_dr1/FWC_dr1/}{https://erosita.mpe.mpg.de/dr1/AllSkySurveyData\_dr1/FWC\_dr1/}} \citep{Mike_2023}. The \texttt{PIB} component includes two power laws to account for the detector noise increase at the low energy, a combination of a power law and an exponential cut-off to model the signal above $\sim\!1~\mathrm{keV}$, and 21 Gaussian lines to model the fluorescence lines produced through the interaction of high-energy particles with detector components.

\begin{figure}
\centering
\includegraphics[width=\columnwidth]{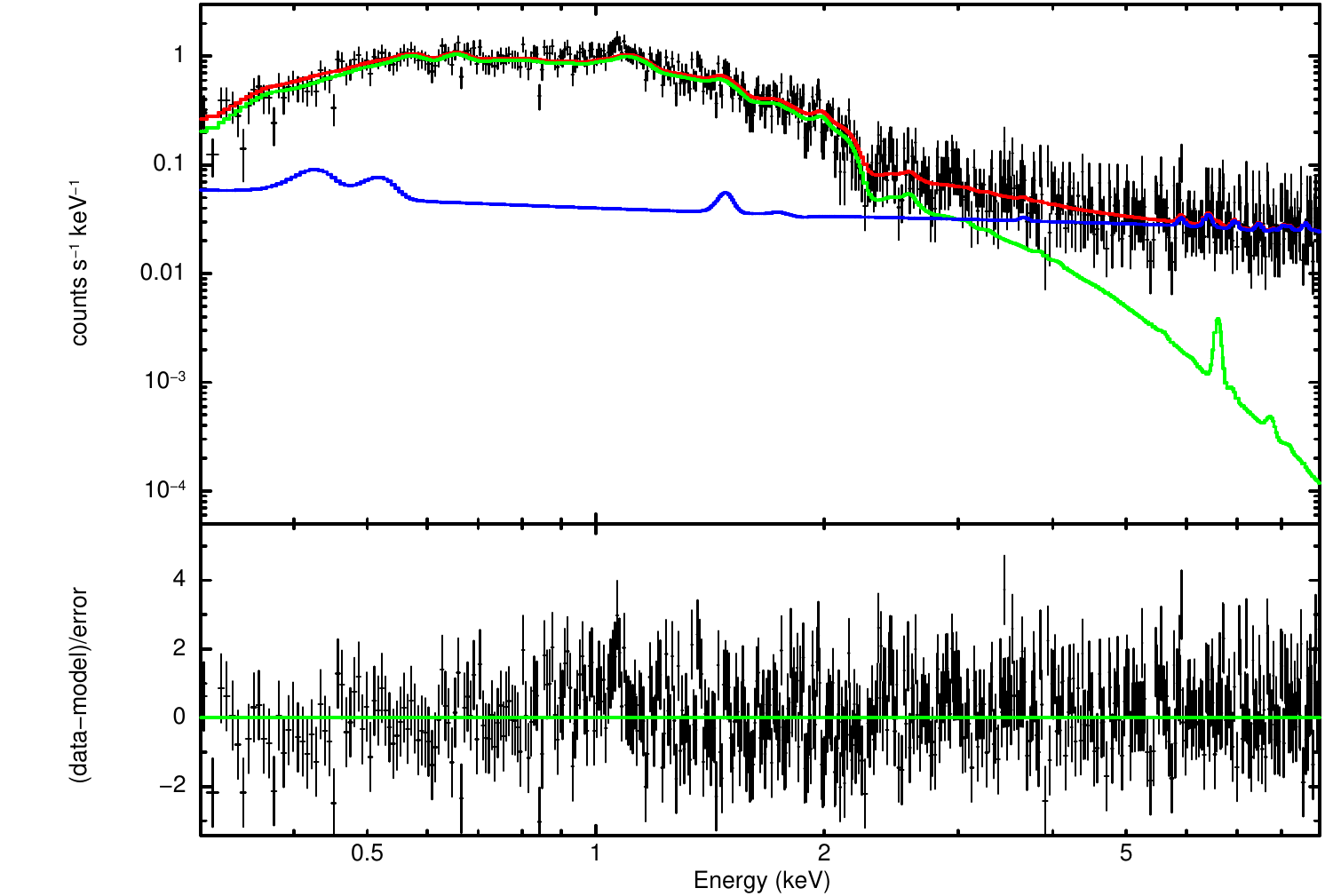}
\caption{TM1 spectrum fitted in the $0.3-9.0~\mathrm{keV}$ energy band (top plot) and the fit residual (bottom). The black data points are the spectral data points extracted from an annulus of $0.2-0.5R_{500}$ around the Centaurus cluster center. The red line is the total model, the green line is the sky component, and the blue line is the PIB component.}
\label{fig:eROSITA_spectrum}
\end{figure}

\par
The fitting procedure is the following: first, the CXB spectra of all TMs were fitted with the first term of Eq.~\ref{eq:spectral_model}, i.e., the CXB components (LHB, MWH, PL, and eROSITA Bubble), to constrain the best-fit values of the normalizations of the CXB components. For all the spectral fitting to acquire the ICM properties, the CXB spectra used were extracted from the magenta boxes shown in Fig.~\ref{fig:conf} (see Sect.~\ref{sec:erobubble}). Afterward, the source and CXB spectra were fitted simultaneously with the complete model. The CXB parameters were linked across the spectra, except for the normalization of the eROSITA Bubble component. The starting parameters of the CXB normalizations were set to be the best-fit values obtained from the first fit. We fixed the eROSITA Bubble normalization of the CXB spectra to the value from the CXB analysis (see Sect.~\ref{sec:erobubble}) and thawed this parameter in the source spectra. All normalizations, including those of the instrumental background components and the CXB components, were freed during the fit. We note that the spatial and temporal variations of the PIB are small \citep{Reiprich_2021}, and we left the PIB normalizations free to allow flexibility in the fitting and take into account the PIB uncertainties. Additional spectral analysis, where the PIB normalizations were frozen, was performed. The results were in good agreement with the free PIB normalizations within $\sim0.6\sigma$ of the statistical errors and therefore, will not change the findings and discussion presented below. The reported results are those of the free PIB normalization version.
\par
The parameters of the source emission represented by \texttt{apec$\mathtt{_4}$}, namely $k_\mathrm{B}T$, $Z$, and $norm$, were left to vary. The redshift was fixed to the Cen 30 redshift, $z=0.0104$ (see Sect.~\ref{sec:ned}). Given that the Centaurus cluster has an average temperature of $3.68\pm0.06~\mathrm{keV}$ \citep{Reiprich_2002}, the fitting was carried out over the full energy range of $0.3-9.0~\mathrm{keV}$ for TM8 and $0.5-9.0~\mathrm{keV}$ for TM9 to accurately assess the hotter regions of the cluster. We adopted the C-statistic \citep{Cash_1979} for estimating the parameters and we used the abundance table from \cite{Asplund_2009}. The analysis was done in effectively ten bins in full azimuthal configuration (annuli) and each sector out to $R_{200}$. Additionally, for full azimuthal configuration, we divided the central $0-2.50'$ area into three annuli, namely $0-0.83'$, $0.83-1.67'$, and $1.67-2.50'$, and we split the outer $R_{500}-R_{200}$ region into two, namely $R_{500}-0.8R_{200}$ and $0.8R_{200}-R_{200}$ bins. For the sectors, we kept these two bins as an individual bin to retain good statistics. The configuration of the sectors and the characteristic radii are displayed in Fig.~\ref{fig:conf}. We show an example of the eROSITA spectrum and its fitted model in Fig.~\ref{fig:eROSITA_spectrum}.

\subsection{The eROSITA Bubbles and cosmic X-ray background}\label{sec:erobubble}
\begin{table}
    \centering
    \caption{CXB information for the spectral analysis of the Centaurus cluster.}
    \resizebox{\columnwidth}{!}{\begin{tabular}{c c c}
    \hline
    \hline
Component & Parameter & Value\\
\hline
\texttt{apec$\mathtt{_1}$} (LHB) & $k_\mathrm{B}T^{456}$ [keV] & $0.115_{-0.002}^{+0.002}$\\
& $Z~[Z_\odot]$ & 1\\
& $z$ & 0\\
& $norm^{*, 456}$ & $(1.42_{-0.05}^{+0.09})\times10^{-6}$\\[5pt]
\hline
\texttt{apec$\mathtt{_2}$} (MWH) &  $k_\mathrm{B}T^{456}$ [keV] & $0.207_{-0.002}^{+0.003}$\\
& $Z~[Z_\odot]$ & 1 \\
& $z$ & 0\\
& $norm^{*, 456}$ & $(1.71\pm0.05)\times10^{-6}$\\[5pt]
\hline
\texttt{apec$\mathtt{_3}$} &  $k_\mathrm{B}T^{1}$ [keV] & $0.20_{-0.0001}^{+0}$\\
(eROSITA Bubbles) & $Z~[Z_\odot]$ & 1 \\
& $z$ & 0\\
& $norm^{*, 456}$ & $<2.80\times10^{-8}$\\[5pt]
\hline
\texttt{powerlaw} & $\Gamma$ & 1.46\\
(unresolved sources) & $norm^{\ddagger, 456}$ & $(6.66_{-0.14}^{+0.13})\times10^{-7}$\\[5pt]
\hline
\multicolumn{3}{l}{\footnotesize $^{456}$ from combined box 4N, 5N, 6N, 4S, 5S, and 6S, $^1$ from combined box 1N and 1S}\\
\multicolumn{3}{l}{\footnotesize $^\dagger$[10$^{22}$ atoms cm$^{-2}$], $^*[\mathrm{cm}^{-5}/\mathrm{arcmin}^2]$, $^{\ddagger}$[photons/keV/cm$^2$/s/arcmin$^2$ at 1 keV]}\\
    \hline
    \hline
    \end{tabular}}
    \label{tab:sky_BG}
\end{table}

ROSAT All-Sky Survey (RASS) discovered a soft X-ray emission that is a part of the North Polar Spur and Loop I \citep{Egger_1995}. The first eROSITA All-Sky Survey (eRASS:1) image further reveals a huge structure in the southern sky analogous to the northern structure \citep{Predehl_eROBub}. These large X-ray structures form a pair of `bubbles' and are dubbed as the `eROSITA Bubbles' \citep{Predehl_eROBub}. The eROSITA Bubbles appear as annuli or shells centered about the Galactic center with extensions of about $80^\circ$ in longitude and $85^\circ$ in latitude. Although larger, the eROSITA Bubbles show great similarity in morphology to the Fermi Bubbles \citep{Su_2010_Fermi} and are therefore thought to originate from the same source, large energy injections from the Galactic center \citep{Predehl_eROBub}.
\par
Our FoV is partially contaminated by the eROSITA Bubble (bright yellowish emission spreading from the center top to bottom left of Fig.~\ref{fig:rgb}). As can be seen in Fig.~\ref{fig:conf}, the eROSITA Bubbles emission is also inside the $2R_{200}$ of the Centaurus cluster, mainly in the northern and eastern sectors. Due to this, we performed an exhaustive investigation of various regions before settling on a specific (CXB) region for the science analyses. We distributed six pairs of boxes, spreading from north to west ($1-6$N in Fig.~\ref{fig:conf}) and from east to south ($1-6$S) outside the $2R_{200}$ to minimize the emission from the cluster. Each box has the size of $1^\circ\times1^\circ$.
\par
First, we performed a spectral fitting in these boxes with a typical CXB model that includes the foreground emission from LHB and the MWH, as well as the emission from the unresolved sources. The LHB and MWH are represented with an \texttt{apec} component each. In this fit, we fixed the temperatures of the LHB to MWH to 0.1 keV and 0.25 keV, whereas their metal abundances and redshift are fixed to $1Z_\odot$ and 0, respectively. Based on this analysis, we observe that the soft band flux ($f_\mathrm{X,0.5-2~keV}$) decreases with a factor of $1.3-1.7$ outwards and flattens out at box 4N, 5N, 6N, 4S, 5S, and 6S (box 456; magenta boxes). We then used these six boxes to constrain the temperatures of the LHB and MWH and their normalizations, as well as the normalization of the unresolved sources (see Table~\ref{tab:sky_BG}). We found a temperature of $0.115\pm0.002~\mathrm{keV}$ for LHB, which is in good agreement with the RASS result \citep[$ 0.097\pm0.013\pm0.006~\mathrm{keV}$;][]{Liu_2016}. For the MWH, our estimate ($0.207_{-0.002}^{+0.003}~\mathrm{keV}$) agrees within the errors of what was measured from ROSAT spectrum \citep[$0.24_{-0.03}^{+0.08}~\mathrm{keV}$;][]{KuntzSnowden_2000} and \xmm EPIC spectra \citep[$0.204\pm{0.009}~\mathrm{keV}$;][]{Lumb_2002}. It is also broadly in agreement with other eROSITA MWH measurements \citep[$0.15-0.20~\mathrm{keV}$;][]{Ponti_2023, Zheng_2024}.
\par
Afterwards, we fit boxes 1N and 1S (box 1) together with an additional absorbed thermal emission that represents the eROSITA Bubbles (\texttt{apec$\mathtt{_3}$}). Similar to the other foreground components, we fixed the metallicity and redshift to $1Z_\odot$ and 0, respectively. In this step, we fixed the parameters related to the LHB, MWH, and unresolved sources to the best-fit values obtained from box 456 and freed the temperature and normalization of the eROSITA Bubbles. We obtained a best-fit temperature of $0.2~\mathrm{keV}$.
A second \texttt{apec} component was added, however, we could not constrain any additional emission from the Bubbles. Comprehensive eROSITA analyses of the LHB, MWH, and eROSITA Bubbles will be reported in the upcoming papers (Yeung et al.; Ponti et al. in prep.). See also recent soft X-ray background investigation using \suzaku by \cite{Sugiyama_2023}, where they report a $\sim\!0.8~\mathrm{keV}$ component in 56 out of 130 observations. The emission measure of this component is higher toward the Galactic plane, which signifies its Galactic origin. We note that the Centaurus cluster region is outside their studied area.
\par
We listed the results of our local CXB analysis in Table~\ref{tab:sky_BG}. Hence, for the science analyses, we used box 456 as the default CXB region. For instance, in the imaging analysis, the CXB level was calculated to be the average of surface brightness values amongst box 456 ($\mathrm{SB}_\mathrm{CXB}$) and the standard deviation is the root-mean-square deviation of these values ($\sigma_\mathrm{CXB}$) to take into account the spatial variation of the CXB. For the spectral analysis, we froze the temperatures of the CXB components to the values listed in Table~\ref{tab:sky_BG}.

\section{Results}\label{sec:results}
\subsection{Galaxy redshift analysis}\label{sec:ned}
\begin{figure}
\centering
\includegraphics[width=\columnwidth]{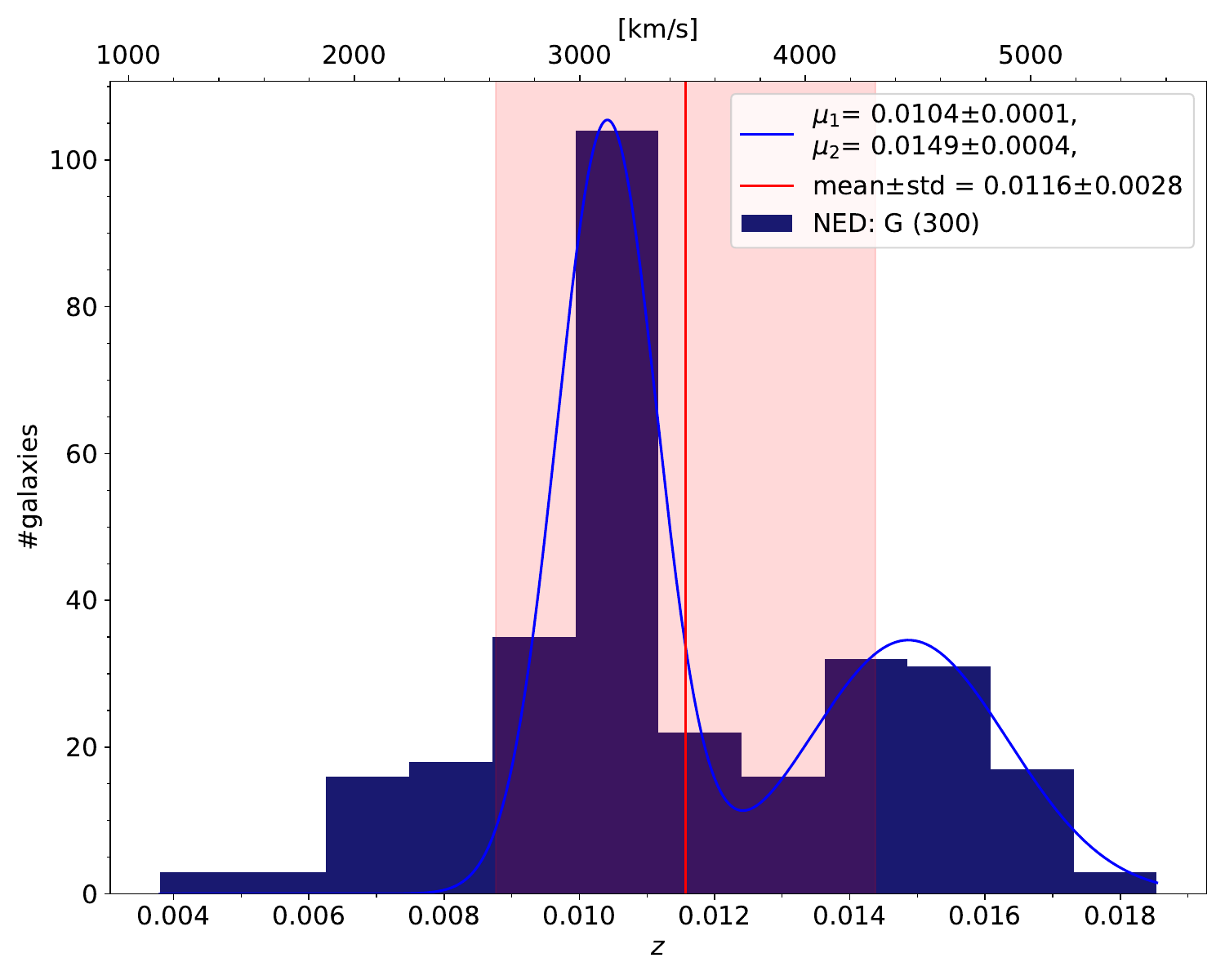}
\caption{NED galaxy redshift distribution within $R_{200}$ of the Centaurus cluster. In total, there are 300 galaxies up to $z=0.02$. The blue line is the double Gaussian fit, the red vertical line and the red shaded area are the mean and the $1\sigma$ standard deviation of the whole sample.}
\label{fig:z-hist}
\end{figure}

\begin{figure*}
\centering
\subfloat[Cen 30 ($0.009<z<0.0118$)]{\includegraphics[width=0.33\textwidth, trim=0.8cm 0.5cm 3.0cm 0.5cm, clip]{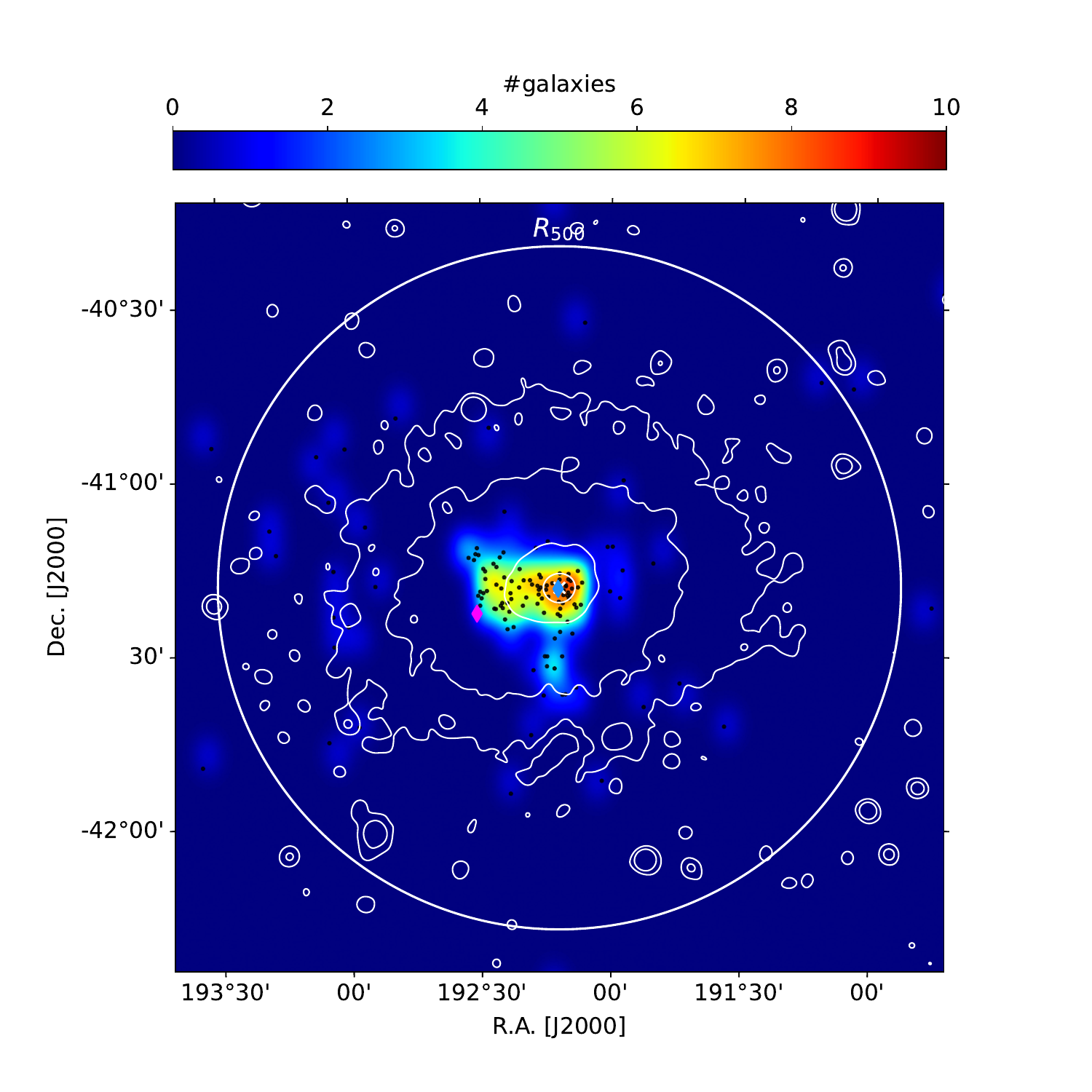}\label{fig:galnumdensa}}
\subfloat[Cen 45 ($0.0119<z<0.0179$)]{\includegraphics[width=0.33\textwidth, trim=0.8cm 0.5cm 3.0cm 0.5cm, clip]{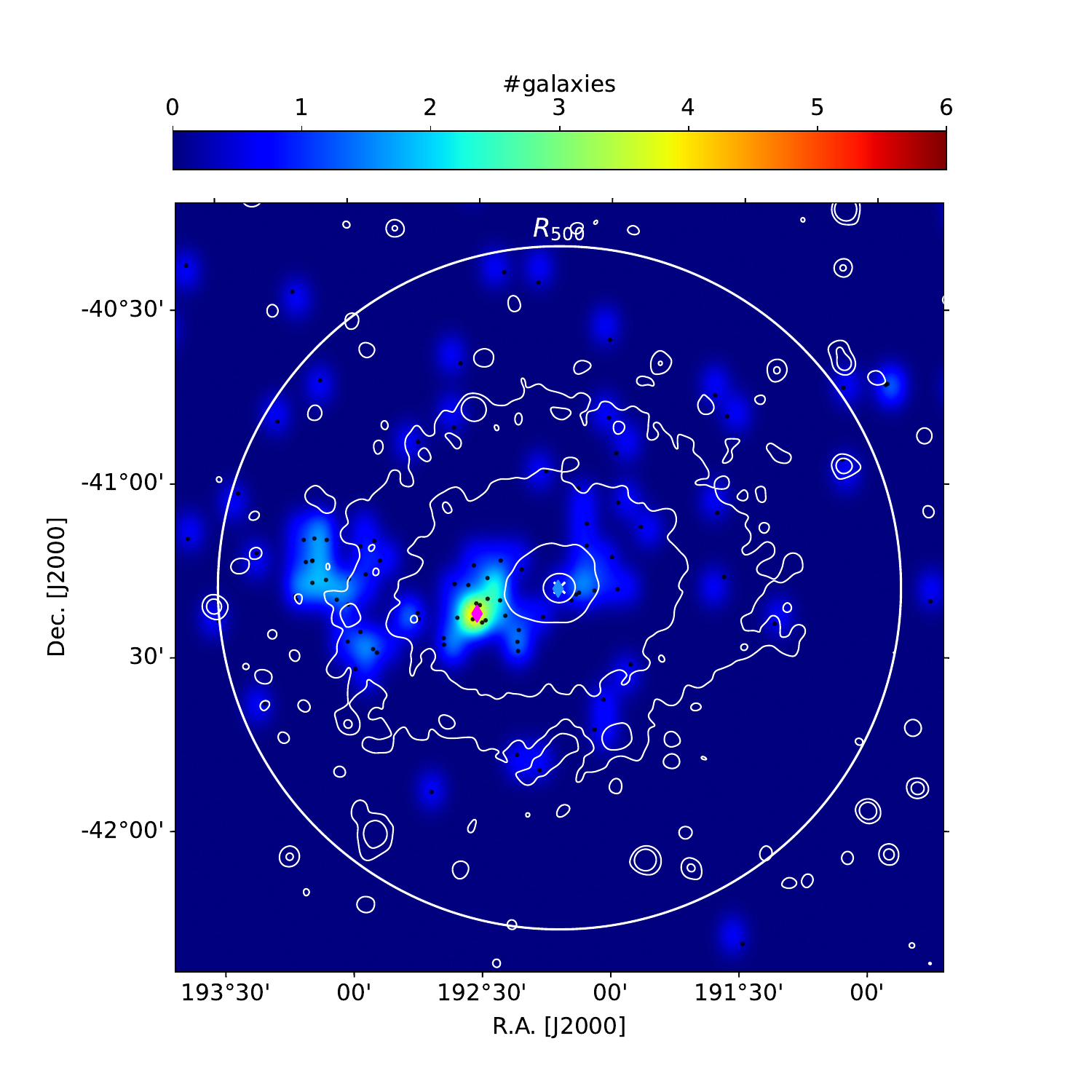}\label{fig:galnumdensb}}
\subfloat[Centaurus ($0.0032<z<0.0200$)]{\includegraphics[width=0.33\textwidth, trim=0.8cm 0.5cm 3.0cm 0.5cm, clip]{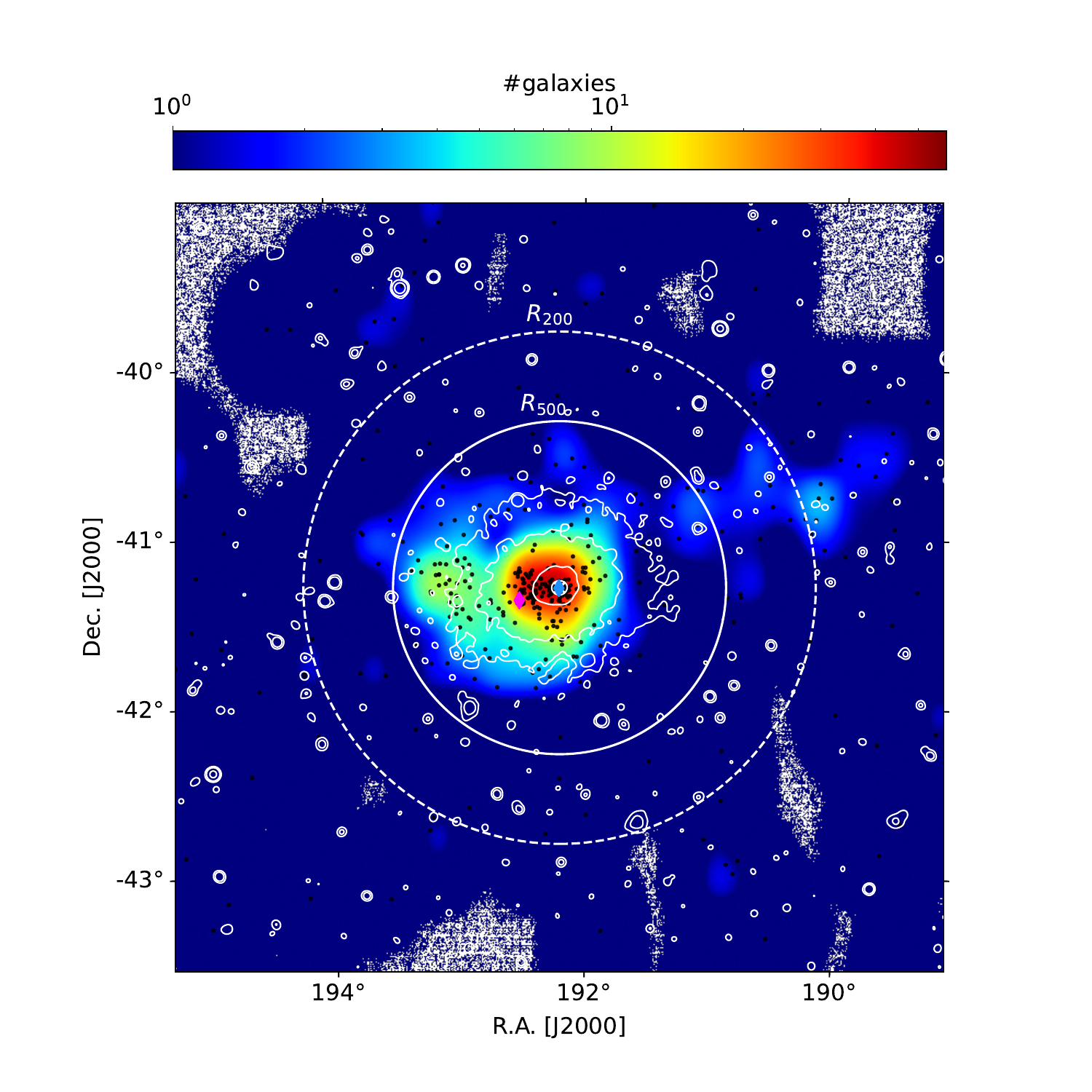}\label{fig:galnumdensc}}
\caption{Galaxy number density maps in different redshift ranges. Blue and magenta diamonds denote the positions of the central galaxies of Cen 30 (NGC 4696 ) and Cen 45 (NGC 4709). The white contours are the eROSITA X-ray contours.}
\label{fig:galnumdens}
\end{figure*}
We compiled a catalog of galaxies with known redshifts within our FoV from NED\footnote{The NASA/IPAC Extragalactic Database (NED) is funded by the National Aeronautics and Space Administration and operated by the California Institute of Technology.}. To inspect the substructures of the cluster, first, we assessed the galaxy redshift distribution within $R_{200}$ (Fig.~\ref{fig:z-hist}). Within this perimeter, we found 300 galaxies with $z\leq0.02$. The mean of the whole sample within the $R_{200}$ is $z=0.0116\pm0.0028$ (red line and red-shaded area in Fig.~\ref{fig:z-hist}), which is in agreement with the value reported in \cite{Piffaretti_2011}. In the distribution, a bimodality is apparent, which has been known to be contributed by the substructures of the Centaurus cluster, which are Cen 30 and Cen 45 \citep{Lucey_1986}. We fitted a double Gaussian function onto the distribution (shown as blue line in Fig.~\ref{fig:z-hist}) and found the peaks at $z_\mathrm{Cen30}=0.0104$ ($v=(3117.8\pm30.0)~\mathrm{km~s^{-1}}$) and $z_\mathrm{Cen45}=0.0149$ ($v=(4466.9\pm119.9)~\mathrm{km~s^{-1}}$).
\par
In Fig.~\ref{fig:galnumdens} we display the galaxy number density maps of three redshift slices for Cen 30 (panel \ref{fig:galnumdensa}; $0.0090<z_\mathrm{Cen30}<0.0118$), Cen 45 (\ref{fig:galnumdensb}; $0.0119<z_\mathrm{Cen45}<0.0179$), and Centaurus cluster as a whole (\ref{fig:galnumdensc}; $0.0032<z<0.020$). The maps were generated by smoothing the spatial bins of their respective galaxy distributions. From the figures, we see that the distribution of the galaxy members of Cen 30 peaks around its central galaxy, NGC 4696 (blue diamond), which also coincides with the cluster center (white cross). The galaxy number density spreads eastward. Similarly, for Cen 45, the peak galaxy number density is situated at the central galaxy NGC 4709 (magenta diamond), $14.9'$ eastward of the cluster center. An excess is also apparent in the east, at the edge of the X-ray contour.
\par
Large-scale filaments can be traced by trails of gas clumps and galaxies \citep[e.g.,][]{Malavasi_2020, Tuominen_2021, Angelinelli_2021}. Therefore, in Fig.~\ref{fig:galnumdensc}, we expanded the view to $\sim\!1.5R_{200}$ and included all galaxies within three times the standard deviation of the mean redshift, which translates to a velocity of $2518.3~\mathrm{km~s^{-1}}$. Within $R_{500}$, a significant galaxy overdensity is apparent in the east. Outside this boundary, a hint of low significance excess is visible in the west, going slightly northward. This feature was referred to as the "Western Branch" in \cite{Lucey_1986} and based on their redshifts, it is composed of the supposed galaxy members of the two substructures. The peak overdensity in this feature, slightly outside of the $R_{200}$, is located at NGC 4603C \citep[$z=0.0104$;][]{Ogando_2008}. We note that these east and west galaxy overdensities match the apparent elongation directions in the gas distribution of the Centaurus cluster.

\subsection{X-ray images}

\begin{figure*}
\centering
\includegraphics[width=\textwidth,trim=0.25cm 0.25cm 0.275cm 0.0cm,clip]{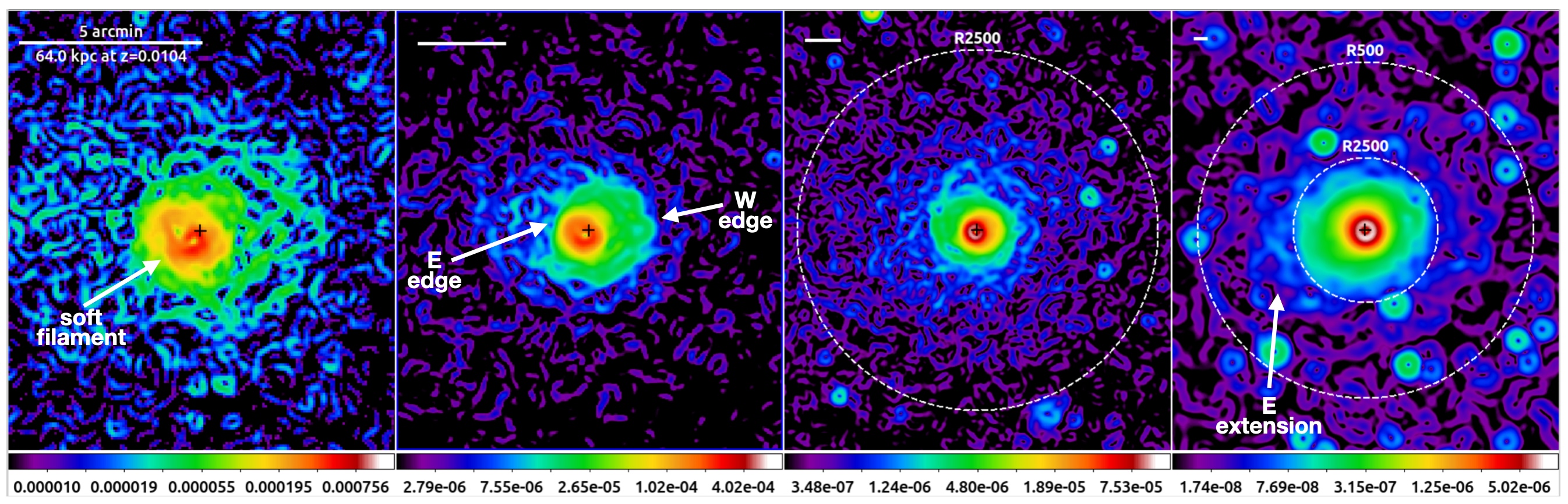}
\caption{GGM-filtered images of the fully-corrected image in $0.2-2.3~\mathrm{keV}$ using scales $\sigma=2,\,4,\,8,\,32$ pixels (1 pixel is $4.0''$). The black cross in the center of each panel marks the cluster center. The white line at the top left of each panel indicates a $5'~(64~\mathrm{kpc})$ length.}
\label{fig:GGM}
\end{figure*}

We present the composite image of the Centaurus cluster combining nine sky tiles of eRASS:5 data in Fig.~\ref{fig:rgb}. The red, green, and blue channels are in the $0.2-0.6~\mathrm{keV}$, $0.6-1.0~\mathrm{keV}$, and $1.0-2.3~\mathrm{keV}$, respectively. Since the lower energy cut of TM9 is $0.8~\mathrm{keV}$, only TM8 was used to produce the red channel image. The image in each channel was PIB-subtracted, exposure-corrected, and Galactic-absorption corrected following the procedure described in Sect.~\ref{sec:drsteps}. To enhance large-scale emission, wavelet filtering was also applied using the same scale for each image. With the same purpose, we generated an adaptively-smoothed image in the $0.2-2.3~\mathrm{keV}$ (Fig.~\ref{fig:conf}).
\par
From Fig.~\ref{fig:rgb}, the emission from the Centaurus cluster is peaked at the center and smooth going outward. The cluster looks undisturbed with a symmetric elliptical shape. Throughout the FoV, many point sources are noticeable. Some with harder spectra appear in white. Overplotted in cyan are the clusters from the eRASS1 catalog in the Western Galactic Hemisphere \citep{Bulbul_2024, Kluge_2024}. The bottom area of the FoV ($b\leq20^\circ$) is excluded from the eRASS1 cluster detection, hence, the absence of clusters is seen in the image. The sizes of the circles correspond to their $R_{500}$. Based on their redshifts (labeled on top of each circle), all of the detected clusters are background objects and thus, excluded from any analyses. A distinctive white extended emission at the top left of the FoV is a known background galaxy cluster, Abell S072.
\par
As mentioned in Sect.~\ref{sec:erobubble}, our FoV is partially contaminated by the eROSITA Bubble emission, which is spreading from center top to bottom left corner of the image and apparent in yellowish (reddish) color in Fig.~\ref{fig:rgb} (\ref{fig:conf}). To investigate at which energy the eROSITA Bubbles roughly diminish in the image, we additionally ran the image correction and adaptive-smoothing procedure in three sub-bands. These images are shown in Fig.~\ref{fig:erobub} left ($0.8-2.3~\mathrm{keV}$), middle ($0.9-2.3~\mathrm{keV}$), and right ($1.0-2.3~\mathrm{keV}$) panels. The surface brightness intensity of the eROSITA Bubble emission decreases toward higher low energy bounds. Based on these sub-band images, we therefore calculated surface brightness profiles using the $1.0-2.3~\mathrm{keV}$ energy band to ensure minimum eROSITA Bubble contamination (see Sect.~\ref{sec:surb}). An attempt was made to spatially subtract the CXB emission (including the eROSITA Bubbles component) from the image. The details of this effort is described in Appendix~\ref{app:erobub}.
\par
Additionally, we applied Gaussian gradient magnitude (GGM) filtering to the $0.2-2.3~\mathrm{keV}$ fully-corrected image. The GGM filtering calculates the magnitude of the surface brightness gradient assuming Gaussian derivatives, which as a result, emphasize features (e.g., edges) in the image \citep{Sanders_2016}. Due to this, the technique is often used to inspect features in clusters, both in simulations and observations \citep[e.g.,][]{Sanders_2016, Sanders_2016edges, Walker_2016}. The GGM filtering is adjusted with the width $\sigma$, where a smaller scale is used to amplify features in the regions with more counts (cluster center) and a larger scale is used for regions with fewer counts (outskirts). We performed the GGM filtering using the available implementation from \textsc{SciPy}\footnote{\href{https://scipy.org/}{https://scipy.org/}} \citep{Scipy_2020}. The resulting GGM-filtered images using $\sigma=2,\,4,\,8,\,32$ are shown in Fig.~\ref{fig:GGM} (left to right). Some features in the central region of the Centaurus cluster are known and well-studied using previous dedicated \chandra and \xmm observations \citep[e.g.,][]{SandersFabian_2002, Fabian_2005, Sanders_2008, Walker_2013, Sanders_2016}, for example, the soft filament in the core that extends toward the northeast seen in the 2-pixel GGM-filtered image, as well as the eastern and western edges in the 4-pixel GGM-filtered image ($\sim2.1'$ and $3.6'$ from the cluster center, respectively). With eRASS:5 data, we are able to examine large-scale features beyond $R_{2500}$. In the 8 and 32-pixel GGM-filtered images, we observe the east-west ICM emission with longer extension toward the east. This eastern extension coincides with galaxy overdensity (see Sect.~\ref{sec:ned}) and the excess will be quantified in Sect.~\ref{sec:surb}.

\subsection{X-ray surface brightness profiles}\label{sec:surb}
\begin{figure*}
\centering
\subfloat[$0\leq r\leq R_{500}$]{\includegraphics[width=0.475\textwidth,trim=0.1cm 0.2cm 0.5cm 0.3cm,clip]{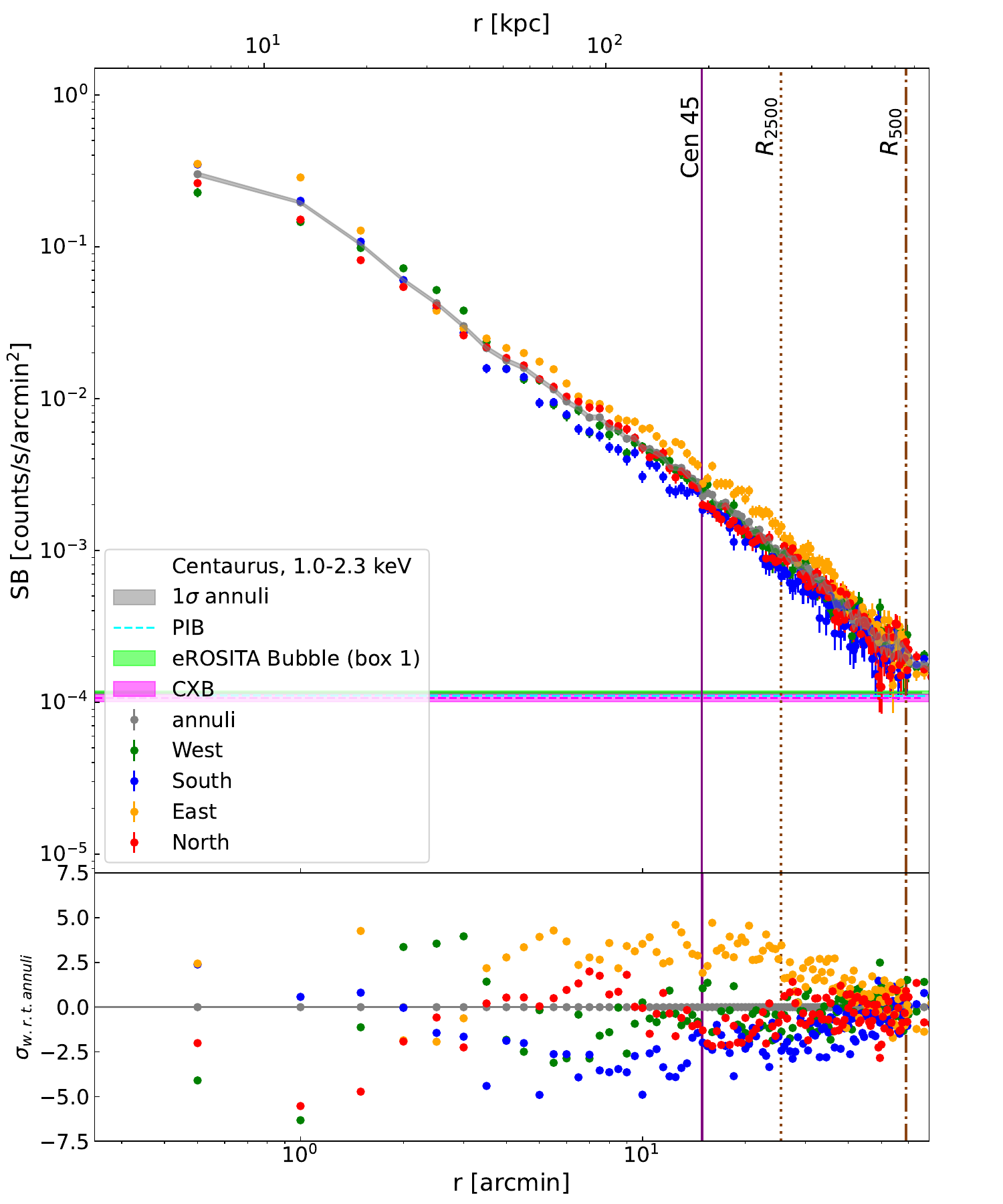}\label{fig:SBprofiles_inhi}}
\subfloat[$R_{500}\leq r \leq 2R_{200}$]{\includegraphics[width=0.475\textwidth,trim=0.1cm 0.25cm 0.5cm 0.3cm,clip]{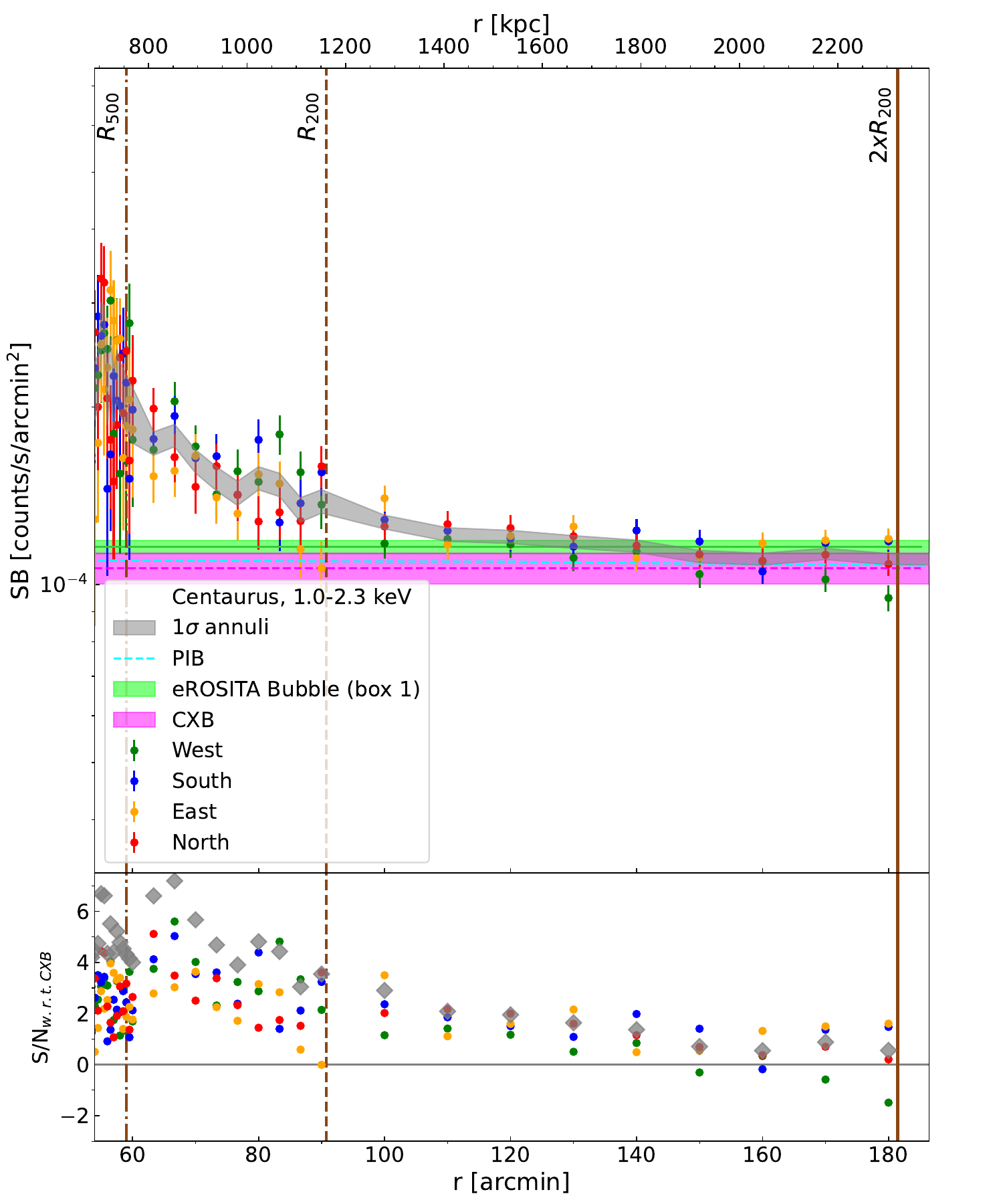}\label{fig:SBprofiles_outhi}}
\caption{PIB-subtracted surface brightness profiles of the Centaurus cluster in the $1.0-2.3~\mathrm{keV}$ band are shown in the top panel in each plot. The different sectors are indicated with different colors. The CXB level and its $1\sigma$ standard deviation are shown as the magenta dashed lines and shaded areas, while the PIB levels are indicated by the cyan dashed lines. The eROSITA Bubble emission calculated from box 1 (see Fig.~\ref{fig:conf}) is marked by the green shaded area. The bottom panels show the significance with respect to the full azimuthal surface brightness profile ($\sigma_\mathrm{w.r.t.annuli}$) and the CXB level ($\mathrm{S/N_{w.r.t.CXB}}$).}
\label{fig:SBprofiles_hi}
\end{figure*}

\begin{figure}[h!]
\centering
\includegraphics[width=\columnwidth]{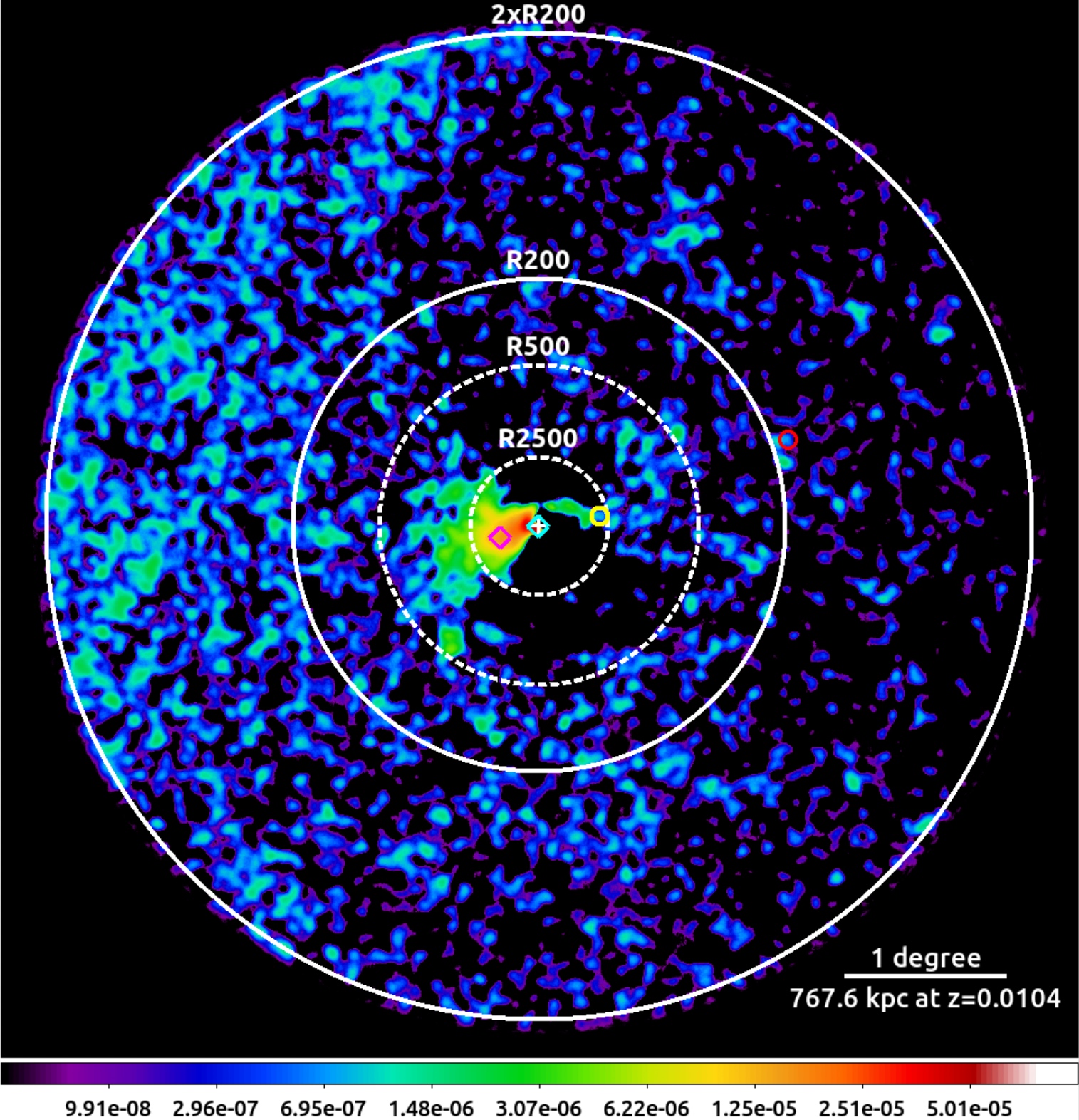}
\caption{Residual image out to $2R_{200}$ in the $0.2-2.3~\mathrm{keV}$ band. The image is Gaussian-smoothed with $\sigma=25$ pixels. The cyan and magenta diamonds denote the positions of NGC 4696 and NGC 4709, while the yellow and red circles denote the positions of NGC 4696B and NGC 4603C.}
\label{fig:residual}
\end{figure}

We present the PIB-subtracted surface brightness profiles of the Centaurus cluster out to $2R_{200}$ in the $1.0-2.3~\mathrm{keV}$ band in Fig.~\ref{fig:SBprofiles_hi}, and in the $0.2-2.3~\mathrm{keV}$ band in Fig.~\ref{fig:SBprofiles}. In each plot, the full azimuthal (annuli) surface brightness values with the $1\sigma$ statistical error are shown as the grey-shaded area. The green, blue, orange, and red data points belong to the western, southern, eastern, and northern sectors, respectively. The configuration of these sectors is shown in Fig.~\ref{fig:conf}. We mark the position of Cen 45, which is located about $15'$ eastward of the center, with a purple vertical line. The CXB level ($\mathrm{SB}_\mathrm{CXB}$) is shown as magenta horizontal-dashed lines and its standard deviation ($\sigma_\mathrm{CXB}$; see Sect.~\ref{sec:erobubble}) is as magenta shaded area. The PIB profiles, calculated from the full azimuthal configuration, are represented by the cyan dashed lines in all of the plots. We also estimated the surface brightness of box 1 (see Sect.~\ref{sec:erobubble} and Fig.~\ref{fig:conf}) as an approximate of the maximum level of the eROSITA Bubble emission in the field. The eROSITA Bubble levels in the two energy bands are displayed as the green shaded area. In the $1.0-2.3~\mathrm{keV}$ band, the eROSITA Bubble level roughly coincides with the CXB level, signifying there is little to no contribution from this additional foreground emission in this energy band. Using the full azimuthal surface brightness profile calculated in $0.2-2.3~\mathrm{keV}$ band, we also constructed a spherical surface brightness image out to $2R_{200}$, which was subtracted from the fully corrected image in the corresponding band to yield a residual image. The Gaussian-smoothed ($\sigma=25~\mathrm{pixels}$) residual image is displayed in Fig.~\ref{fig:residual}.

\subsubsection{Inner ($r\leq R_{500}$)}
In Fig.~\ref{fig:SBprofiles_inhi} and \ref{fig:SBprofiles_infull}, we show the surface brightness profiles out to $R_{500}$ in the energy band of $1.0-2.3~\mathrm{keV}$ and $0.2-2.3~\mathrm{keV}$ (top plot) and the significance deviation profile of each sector with respect to the full azimuthal surface brightness profile ($\sigma_\mathrm{w.r.t.annuli}$; bottom plot). Comparing the profiles between the two energy bands, we observe only minor differences, such as lower surface brightness amplitudes of the profiles and the CXB level, suggesting that there is little to no impact from the eROSITA Bubbles for inner $r<R_{500}$ regime. We keep the axis scale consistent for these inner surface brightness profiles to show the mentioned differences. The significances described in the following were calculated from the $0.2-2.3~\mathrm{keV}$ profiles.
\par
The western and eastern excesses, as well as the southern depression, are still visible. In the west at around $2-3'$, an excess is observed with an average significance of $3.8\sigma$. The significance drops to $1.3\sigma$ in the next bin. The location of the western excess matches the location of the western edge that is apparent in the 2 and 4-pixel GGM-filtered images (see Fig.~\ref{fig:GGM}).
\par
A striking excess bump of about $3-5\sigma$ is also seen in the eastern sector at $4-25'$. This enhancement coincides with the interaction location between Cen 30 and Cen 45, which is shown as the galaxy overdensity in Fig.~\ref{fig:galnumdens}. From the eastern surface brightness profile in the $0.2-2.3~\mathrm{keV}$, the average significance surface brightness excess with respect to the average surface brightness values in different segments are $4.0\sigma$ for $0-14.8'$, $4.2\sigma$ for $14.8-25.3'$, and $2.5\sigma$ for $25.3-35.3'$. Outside the core, the upper limits of these segments correspond to roughly the position of the NGC 4709 galaxy (Cen 45), $R_{2500}$, and $0.6R_{500}$ from the cluster center, respectively. As a comparison, the significance values in the opposite western sector are $-1.4\sigma$, $-0.7\sigma$, and $-0.2\sigma$ for each segment. Moreover, we observe surface brightness deficits of $3-5\sigma$ in the southern sector at $5-14'$. These features are further emphasized in the residual image (Fig.~\ref{fig:residual}).

\subsubsection{Outer ($R_{500}<r<2R_{200}$)}
Similarly, the top plots of Fig.~\ref{fig:SBprofiles_outhi} and \ref{fig:SBprofiles_outfull} show the surface brightness profiles in the $1.0-2.3~\mathrm{keV}$ and $0.2-2.3~\mathrm{keV}$ bands, but at radial distance between $R_{500}$ and $2R_{200}$. The bottom plots, instead, show the significance level of the source surface brightness with respect to the CXB level (S/N$_\mathrm{w.r.t.CXB}$) and is given as:
\begin{equation}
    \mathrm{S/N}_{\mathrm{w.r.t.CXB}} = \frac{\mathrm{SB}_{r>R_{500}} - \mathrm{SB}_{\mathrm{CXB}}}{\sqrt{\sigma_{r>R_{500}}^2 + \sigma_{\mathrm{CXB}}^2}},
\label{eq:snr}
\end{equation}
where $\mathrm{SB}_{r}$ is the surface brightness value of the bin at $r>R_{500}$ regime, $\sigma_{r}$ is its statistical uncertainties, while $\mathrm{SB}_{\mathrm{CXB}}$ and $\sigma_{\mathrm{CXB}}$ are the sky background surface brightness and its standard deviation.
In the $R_{500}<r<R_{200}$ regime, we observe significant cluster emission, as shown in the $0.2-2.3~\mathrm{keV}$ significance profile for the full azimuthal region (grey diamonds) with significance values spreading between 3.3 and $6.7\sigma$. Between $R_{200}$ and $2R_{200}$, the full azimuthal and northern (red) significance profiles show similar trend of steady increase. The significance values at $2R_{200}$ of the full azimuthal and northern profiles are 3.4 and $5.6\sigma$, respectively. The Eastern (orange) significance profile at this regime shows a similar trend but with higher amplitude. The minimum significance is $3.6\sigma$ at $110'$ and the maximum is $7.1\sigma$ at $190'$. For the southern (blue) profile, excesses are seen, however with less significance in comparison to the northern and eastern sectors. The significance values of the southern sector in $R_{500}<r<R_{200}$ regime are between 1.7 at $160'$ and $5.4\sigma$ at $70'$. Unlike the other sectors, at $r>R_{200}$, the surface brightness profile of the green sector stays around the $1\sigma$ of the CXB level (magenta-shaded area). The trends seen in the profiles of the various sectors in this $0.2-2.3~\mathrm{keV}$ energy band show the levels of eROSITA Bubbles contamination, such that the contamination is the strongest in the east and north, mild in the south, and none in the west. This agrees with the visual inspection (see Fig.~\ref{fig:residual}).
\par
Between $R_{500}$ and $R_{200}$ in the $1.0-2.3~\mathrm{keV}$ full azimuthal surface brightness profile (Fig.~\ref{fig:SBprofiles_outhi}), we observe similarity in the shape of the profile to those of $0.2-2.3~\mathrm{keV}$ profile. For the eastern sector, the average significance value between $60'\leq r \leq 83'$ is $2.6\sigma$. Near the $R_{200}$ mark, the values drop to the CXB level and the bin afterwards jumps to $3.5\sigma$. The significance values of the following eastern bins decrease. As for the other sectors, we see the significance values approach the CXB level between $R_{500}$ and $2R_{200}$. The significance values for the full azimuthal annuli at $R_{500}<r<R_{200}$ regime vary between $3.0\sigma$ and $7.2\sigma$. At $90'~(\sim\!R_{200})$, a $3.5\sigma$ signal above the CXB was detected, then followed by $2.9\sigma$ at $100'~(\sim\!1.1R_{200})$. The smooth significance decrease hits the $1\sigma$ level around $140'~(\sim\!1.5R_{200})$.

\subsection{Spectral analysis}\label{sec:spectro_results}
\begin{figure*}[h!]
\centering
\includegraphics[width=0.49\textwidth,trim=0cm 0.5cm 1.5cm 0.5cm,clip]{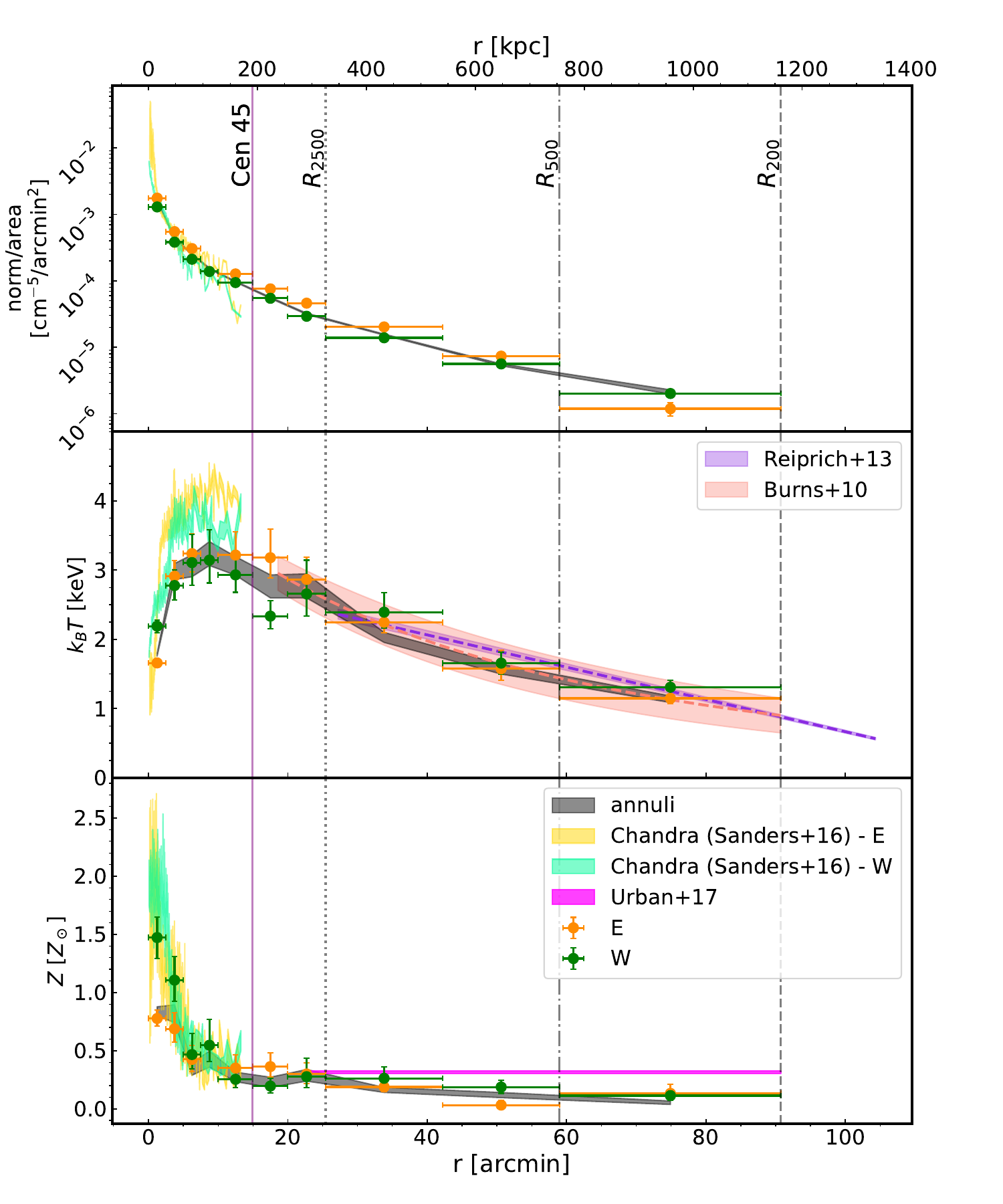}
\includegraphics[width=0.49\textwidth,trim=0cm 0.5cm 1.5cm 0.5cm,clip]{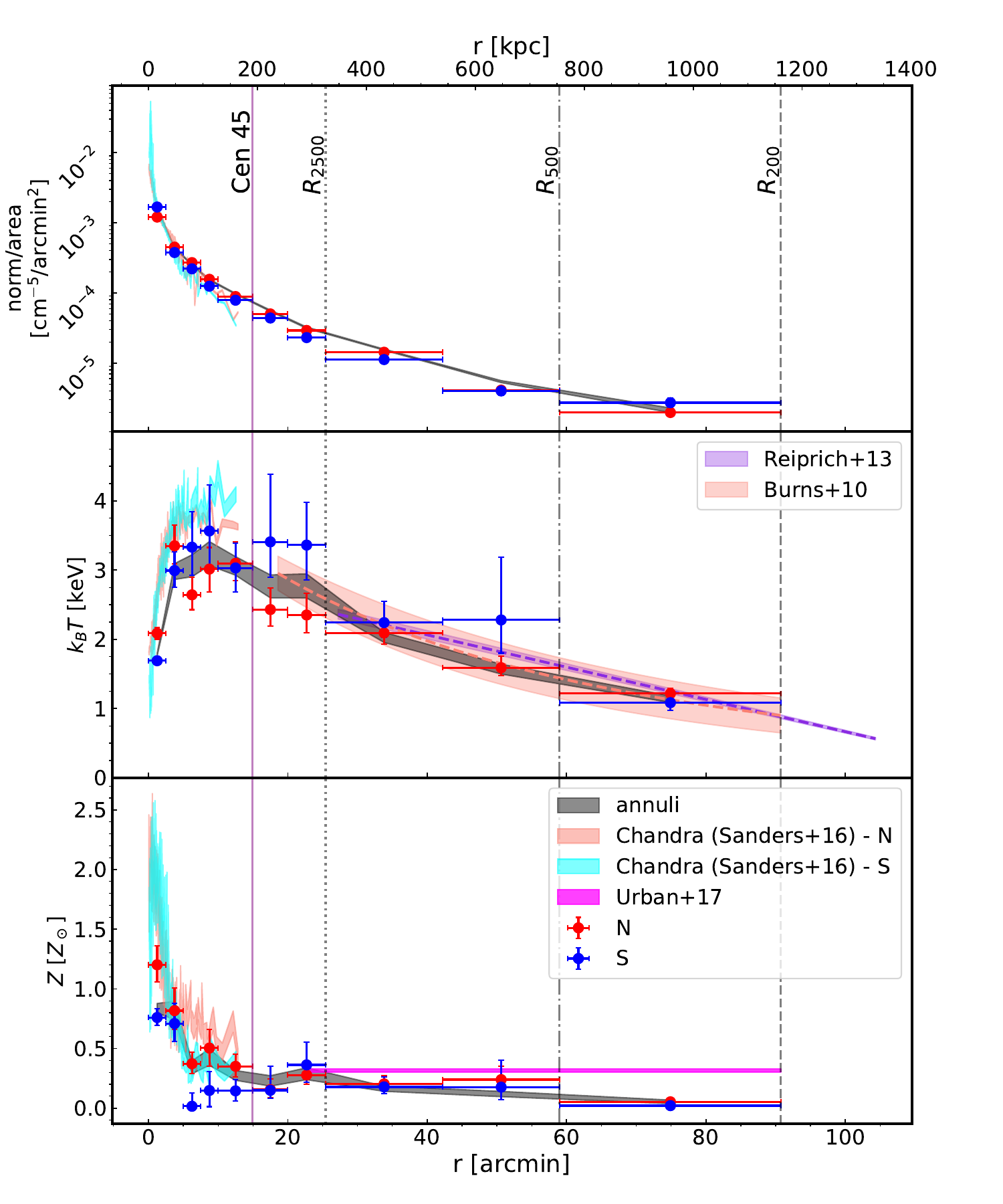}
\caption{The spectral parameter profiles of the Centaurus cluster constrained from eRASS:5 data. \textit{Left:} Eastern (orange data points) and western (green) profiles. \textit{Right:} Northern (red) and southern (blue) profiles. In each panel we show the normalization per unit area (top), gas temperature (middle), and metallicity (bottom) plots. The full azimuthal profile of each parameter is shown as the grey-shaded area. In the temperature panels (middle), we plot the \suzaku cluster temperature fit \citep[purple-shaded area;][]{Reiprich_2013} and the average two-dimensional cluster profile from simulations \citep[pink-shaded area;][]{Burns_2010}. In the metallicity panels (bottom), the average metallicity measurements reported by \cite{Urban_2017} for $r>0.25R_{200}$, $Z_{\mathrm{Fe}}=0.316\pm0.012~\mathrm{Solar}$, is indicated by the magenta shaded area.}
\label{fig:specprofiles}
\end{figure*}

We present the spectral parameter profiles of the Centaurus cluster out to $R_{200}$ obtained from eRASS:5 data in Fig.~\ref{fig:specprofiles}, which are the first detailed profiles of the cluster at $r>30'$. We display the profiles of the eastern (orange data points) and western (green) sectors in the left panel, and the profiles of the northern (red) and southern (blue) sectors in the right panel of Fig.~\ref{fig:specprofiles}. In each panel, we show the normalization per unit area (top), gas temperature (middle), and metallicity (bottom). In each plot, the full azimuthal profile of each parameter is shown as the grey-shaded area. The purple vertical line is the position of the center of Cen 45 from the X-ray center (only relevant for the eastern sector). The characterstic radii are denoted with the black vertical-dotted ($R_{2500}$), dashed-dotted ($R_{500}$), and dashed ($R_{200}$) lines. We also list these results in Table~\ref{tab:spectro}.
\par
To compare the resulting outskirts temperatures, we plot in Fig.~\ref{fig:specprofiles} the temperature fit from \cite{Reiprich_2013} as a purple-shaded area and the average temperature profile from simulations given in \cite{Burns_2010} as a pink-shaded area. The temperature profile from \cite{Reiprich_2013} was acquired from fitting 162 \suzaku cluster profiles and given as $k_\mathrm{B}T(r)=(1.19-0.84r/R_{200})\langle k_\mathrm{B}T \rangle$ at $0.3R_{200}<r<1.15R_{200}$. The profile from \cite{Burns_2010} is the average temperature two-dimensional profile of the simulated clusters using the \textit{Enzo} \textit{N}-body plus hydrodynamics cosmology code \citep{Bryan_2014}. The form for the \cite{Burns_2010} temperature profile is given in their Eq.~8, which we re-arranged into:
\begin{equation}
    k_\mathrm{B}T(r) = A\cdot\langle k_\mathrm{B}T \rangle \left[1 + B \left( \frac{r}{R_{200}} \right)\right]^{\alpha},
\end{equation}
where $A=1.74\pm0.03$, $B=0.64\pm0.10$, and $\alpha=-3.2\pm0.4$. In both temperature functions, $\langle k_\mathrm{B}T \rangle$ is given as the eRASS:5 temperature of Centaurus cluster obtained from an annulus of $0.2-0.5R_{500}$, that is $2.64_{-0.07}^{+0.08}~\mathrm{keV}$.
\par
We also compare the eRASS:5 results with \chandra results out to $\sim\!13'$ (colored-shaded areas). The \chandra information is the adaptation of Fig. 3 or Fig. 6 of \cite{Sanders_2016}, which was obtained from \chandra spectral fittings in the $0.5-7.0~\mathrm{keV}$ band. In order to have a close comparison to the eRASS:5 profiles, the bins in the \chandra spectral maps have been split into the sectors defined in this present work. We report the temperatures returned by eROSITA are lower than \chandra measurements by about $25\%$ on average, which is on the same order as the reported value in the cross-calibration work by \cite{Migkas_2024} \citep[see also][for the eROSITA-\chandra temperature comparisons of the SMACS J0723.3--7327 cluster]{Liu_2023}.
\par
To compare with the \chandra metallicity profiles, we need to scale up the metallicity values from \chandra by a factor of 1.48. The factor is to account for the different abundance tables and is calculated by dividing the solar Fe/H ratio from the abundance table assumed in \cite{Sanders_2016} \citep[$4.68\times10^{-5}$]{Angr_1989} with the Fe/H ratio assumed in this work \citep[$3.16\times10^{-5}$]{Asplund_2009}.
\par
Additionally, we fitted the cluster with only atomic hydrogen ($N_\mathrm{HI}$) absorption. We found systematically higher temperature values in comparison to the spectral fitting results using the $N_\mathrm{Htot}$ values. The same outcome is also reported in \cite{Rossetti_2024}. In general, we find that the discrepancies for bins with higher temperatures ($k_\mathrm{B}T>2~\mathrm{keV}$) are higher ($>20\%$). We note that the presented spectral analysis are those constrained with the $N_\mathrm{Htot}$ values. We show comparison plots of the derived spectral parameter profiles using the different absorption values in Appendix~\ref{App:spectro}, Fig.~\ref{fig:nhI-nhtot}.

\subsubsection{Full azimuthal profiles}
We observe a low temperature ($1.43_{-0.01}^{+0.03}~\mathrm{keV}$), metal-rich ($0.67\pm0.05Z_\odot$) gas in the core of Centaurus cluster, as also seen in other X-ray observatories \citep[e.g.,][]{Allen_1994, Sanders_2016, Gatuzz_2022}. Looking at the full azimuthal profile, the temperature increases gradually to $\sim\!3.0~\mathrm{keV}$ at $6.3'$, then it drops from $22.7'$ ($0.9R_{2500}$) outward. From the peak to the outermost bin, the temperature decreases by a factor of $2.8$.
\par
For the metallicity profile, we identified a peak at $1.3'$ of $1.18_{-0.09}^{+0.10}Z_\odot$. From $3.8'$ to $6.3'$, the metallicity drops by a factor of 2 to a sub-solar value of $0.34_{-0.05}^{+0.06}Z_\odot$. Outward, the metallicity decreases smoothly. Near $R_{500}$, we report a metallicity of $0.12\pm0.02Z_\odot$ and a best-fit value of $0.04Z_\odot$ ($R_{500}-0.8R_{200}$) and $0.09Z_\odot$ ($0.8R_{200}-R_{200}$). We note that for consistency and easier comparison with the other sectors, the combined central ($0-2.5'$) and outermost bins ($59.0-90.8'$) of the full azimuthal profile are used.

\subsubsection{Eastern and western profiles}
In agreement with the surface brightness profile and residual image, there is an excess normalization per area with respect to the full azimuthal values in the eastern sector for all bins, except for the outermost one. At the bin before ($10.0-15.0'$) and after the center of Cen 45 ($15-20.0'$), there is a $31\%$ and $36\%$ excess of $5.4\sigma$ in both bins. The normalization excess peaks at $20.0-25.4'$ bin with $45\%$ and $7.7\sigma$, then falls to $30\%~(7.0\sigma)$ and $36\%~(2.6\sigma)$ in the following bins. In the last two bins (from $0.7R_{500}-R_{500}$ to $R_{500}-R_{200}$), the normalization per area drops with a factor of $\sim\!6$, whereas it is only a factor of $\sim\!3$ for the same bins in the west.
\par
The shape of the eastern and western temperature profiles (middle plot of the left panel in Fig.~\ref{fig:specprofiles}) shows a great resemblance to the full azimuthal temperature profile. At the central bin ($0-2.5'$), the western sector shows a higher temperature ($2.19_{-0.10}^{+0.08}~\mathrm{keV}$) in comparison to the eastern sector ($1.66\pm0.03~\mathrm{keV}$) with a significance of $5.2\sigma$. In general, except for the first bins, the temperature data points of these sectors are within $1.6\sigma$ of the full azimuthal values.
\par
In comparison to the first bin of the eastern best-fit metallicity ($0.78Z_\odot$), the value in the west is higher by a factor of 2 ($1.47Z_\odot$). The metallicity profiles of both sectors fall after the first bin and flatten outward.
At larger radii, for instance, at $59.0-90.8'$ ($R_{500}-R_{200}$), the metallicity values are low, that is $Z=0.07-0.16Z_\odot$ for the western and $Z=0.08-0.21Z_\odot$ for the eastern sectors.

\subsubsection{Northern and southern profiles}
Similarly, the profiles of the northern and southern sectors exhibit the low-temperature, high-metallicity gas in the core. Temperature and metallicity asymmetry are also observed for the first bins. For instance, the northern sector shows a higher temperature ($2.09_{-0.09}^{+0.09}~\mathrm{keV}$) than its counterpart bin ($1.69_{-0.03}^{+0.04}~\mathrm{keV}$) with a significance of $4.2\sigma$. The metallicity in the north is $1.2_{-0.15}^{+0.16}Z_\odot$, while it is $0.76_{-0.07}^{+0.07}Z_\odot$ in the south. The northern metallicity in the core is higher than in the south by a factor of $1.6$, with a significance of $2.7\sigma$. Other fluctuation is the southern temperature ($2.28_{-0.48}^{+0.91}~\mathrm{keV}$) excess with respect to its northern ($1.59_{-0.11}^{+0.16}~\mathrm{keV}$) counterpart at the $42.2-59.0'$ bin. However, this temperature excess is only $1.4\sigma$ ($1.3\sigma$) to the northern (full azimuthal) profile. The large error bars, especially of the southern profiles, can be attributed to the fewer statistics that are evident in the imaging analysis (for example, the residual image in Fig.~\ref{fig:residual} and surface brightness profiles in these directions in Fig.~\ref{fig:SBprofiles}). This is also expected as the north-south direction is the minor-axis direction of the cluster. Between 2.5 and $59'$, the average temperature uncertainty is $0.50~\mathrm{keV}$ for the southern sector and $\sim\!0.25~\mathrm{keV}$ for the other sectors.
\par
The metallicities in the second bins are high with $0.82Z_\odot$ for the northern sector and $0.71Z_\odot$ for the southern sector. From the second to the third bin of the southern sector, a dramatic drop of $3.5\sigma$ is observed. From this point to $R_{500}$, the southern metallicity profile appears flat with an average value of $0.19Z_\odot$. For the northern sector, the metallicities at the intermediate radial range of $5.0-15.0'$ are relatively high with an average of $0.44Z_\odot$ and ranging between 0.29 and $0.66Z_\odot$. At $R_{500}-R_{200}$, we report low metallicity ranges of $0.02-0.08Z$ in the north and $0.003-0.05Z_\odot$ in the south.

\subsubsection{Two-temperature profile}\label{sec:spectro_results_2T}
We performed a two-temperature (2T) fitting to the full azimuthal bins up to $2.5'$ from the center. To reduce the degree of freedom during the fit, we linked the metallicities of the two components. The temperature profile of the hotter component resembles closely the single component temperature profile (grey-shaded area), while the temperatures of the cooler component across the bins are relatively flat, ranging between $0.8$ and $1.7~\mathrm{keV}$. We note that based on the Akaike information criterion \citep[AIC;][]{AIC} and the Bayesian information criterion \citep[BIC;][]{BIC}, there is no significant improvement in the fit when adding a second component, except for the first bin ($0-0.8'$) where the 2T fit is a more suitable model to describe the spectra. The results of the 2T-fit are listed in Table~\ref{tab:spectro2T}.

\section{Discussion}\label{sec:discussion}
\subsection{Galaxy optical redshift analysis}
We performed a galaxy redshift analysis in the eRASS:5 Centaurus field utilizing the available photometric redshift compiled from NED (see Sect.~\ref{sec:ned}). We found 300 galaxies within the $R_{200}$ with $z<0.02$. The distribution shows a bimodality (see Fig.~\ref{fig:z-hist}), peaked at $z=0.0104\pm0.0001$ ($v=3117.8\pm30.0~\mathrm{km~s^{-1}}$) and $z=0.0149\pm0.0004$ ($v=4466.9\pm119.9~\mathrm{km~s^{-1}}$), which has been known to represent the Cen 30 and Cen 45 substructures \citep[e.g.,][]{Lucey_1980, Lucey_1986}. The spatial distribution of the galaxy members of each structure is shown in Fig.~\ref{fig:galnumdensa} (Cen 30) and Fig.~\ref{fig:galnumdensb} (Cen 45). In these panels, the peaks of the spatial distributions match well with the position of the brightest galaxies of the components, that is NGC 4696 for Cen 30 (blue diamond) at the very center of the cluster and NGC 4709 for Cen 45 (magenta diamond) at $\sim\!14.9'$ eastward of the center. Despite the $\sim\!1500~\mathrm{km~s^{-1}}$ line-of-sight velocity separation, \cite{Lucey_1986} argued that Cen 30 and Cen 45 belong to the same cluster, which was concluded after comparisons of the $U-V$ color-magnitude relations, luminosity functions, and galactic radius distributions of both substructures.
\par
Meanwhile, the redshifts inferred through the ICM velocities using deeper X-ray observations (e.g., \xmm and \suzaku) at the locations of these substructures suggest that both Cen 30 and Cen 45 are at $z\approx0.0104$ \citep{OtaYoshida_2016, Gatuzz_2022}. The reported radial velocity relative to the optical redshift (line-of-sight bulk velocity) for Cen 45 is $<760~\mathrm{km~s^{-1}}$ ($90\%$ limit) from \suzaku spectra \citep{OtaYoshida_2016} and $216\pm258~\mathrm{km~s^{-1}}$ from \xmm spectra \citep{Gatuzz_2022}, smaller than the optical estimate of $\sim\!1500~\mathrm{km~s^{-1}}$. Such offset between the ICM mass centroids and the galaxy distribution along the line of sight is a hint of previous subcluster merger \citep{OtaYoshida_2016}, as has previously been observed in other on-going merging systems, for example, the Bullet cluster, \citep{Markevitch_2004, Clowe_2006} and A2744, \citep{Merten_2011}.
\par
In the galaxy distribution map shown with wider FoV (Fig.~\ref{fig:galnumdensc}), we observed a stripe of low significant galaxy overdensity stretching westward. This structure is known as the Western Branch \citep{Lucey_1986, LuceyIII_1986} and is composed of the galaxy members of both structures (based on their redshifts). A small clump in this stripe is seen slightly outside the $R_{200}$ and centered at the NGC 4603C galaxy. However, unlike other galaxy distribution features, we do not observe any correlation of the Western Branch with X-ray emission. Given the separation of the Western Branch from the main location of the distribution of the Cen 45 member galaxies in the east, the galaxies with similar redshifts in the Western Branch are likely background galaxies, decreasing further the significance of this structure.

\subsection{X-ray analysis}
\subsubsection{Nucleus}
In the GGM images (Fig.~\ref{fig:GGM}), we observed some well-known features in the core and also emphasized large-scale features beyond $R_{2500}$. For instance, the soft plume-like filament originating from the core that extends toward the northeast can be clearly seen in the 2-pixel GGM-filtered image. This central region has been studied in great detail using \chandra observations with better spatial resolution \citep[e.g.,][]{SandersFabian_2002, Fabian_2005, Sanders_2016}. As NGC 4696 hosts a low-power radio source, PKS $1246-410$, radio data were also employed \citep{Taylor_2002, Taylor_2006}. The multi-wavelength studies using \chandra X-ray observations and VLA and VLBA radio observations reveal a wealth of structures in the nucleus of the Centaurus cluster. For example, it was found that the plume is the most prominent below $\sim1~\mathrm{keV}$ and it consists of three soft filaments. The nucleus, where the plume seems to originate, coincides with optical filaments and dust lane from NGC 4696 \citep{Fabian_1982, Sparks_1989, Hamer_2019}. On the sides of the nucleus are cavities that anti-correlate with radio lobes. These authors concluded that the plume and other features in the core of the Centaurus cluster, including cavities, edges, shocks, bay, as well as core metallicity asymmetries, are the result of AGN feedback from the NGC 4696 and sloshing motion. We refer to the mentioned previous works for further details of this structure.
\par
The trend shown in the ICM properties (normalization per area, temperature, and metallicity) is similar across the different region configurations (Fig.~\ref{fig:specprofiles}), with some variations that highlight the features which will be discussed in the following subsections.
\par
\cite{Fukushima_2022} conducted a detailed study on the chemical enrichment in the core of the Centaurus cluster, utilizing deep data from both \chandra and \xmm (MOS and RGS). Their findings reveal flat metallicity profiles towards the cluster center, alongside inconsistencies between different detectors and plasma models. Similarly, to inspect the core metallicity drop seen in eRASS:5 data, we fitted the spectra of the innermost full azimuthal bins ($0-150''$) with the SPEX Atomic Code and Tables \citep[SPEXACT;][]{Kaastra_1996_Spex} version 3.05.00 \citep{Kaastra_2018_Spex305}. We show the core metallicity profiles and the comparison plot using the two atomic databases in Appendix~\ref{App:spectro}, Fig.~\ref{fig:spex} left and right, respectively. The metallicities derived using SPEXACT are higher than those from AtomDB \citep{Smith_2001, Foster_2012_atomdb} version 3.0.9, with a relative difference of about $39\%$, $22\%$, and $16\%$ for the first, second, and third bins, respectively. The use of SPEXACT increases the amplitude of the metallicities in these bins, indicating systematic uncertainties in atomic databases \citep{Fukushima_2022}. Nevertheless, the observed drop in metallicity remains evident regardless the used atomic databases.
\par
To investigate the multi-temperature structure effect, we performed a two-temperature spectral fitting to the bins at $r<2.5'$ (2T; see Sect.~\ref{sec:spectro_results_2T}). Based on the AIC and BIC statistical tests, we found that a 2T model only statistically improves the fit of the innermost bin ($0-0.8'$). The lowest best-fit temperature value decreases from $1.43~\mathrm{keV}$ (1T) to $1.0~\mathrm{keV}$ (2T). \cite{Sanders_2016} attribute the multi-phase nature of the ICM in the core ($r<10~\mathrm{kpc}$) to the plume structure, and not a result of the projection effects from the temperature gradient, which is in accordance with our findings\footnote{$0.8'$ is $\sim\!10.7~\mathrm{kpc}$ at $z=0.0104$}. A similar conclusion was drawn in \cite{Ikebe_1999} using \rosat PSPC and ASCA data. Moreover, we acquired a metallicity value of $1.59_{-0.18}^{+0.22}Z_\odot$, which is around $2.4$ times higher than the 1T value ($0.67_{-0.05}^{+0.05}Z_\odot$) with a significance of $4.9\sigma$, removing the central metallicity drop seen in the 1T profile. \cite{Buote_2000} investigated the metallicities in the core of 12 bright galaxy groups with \textit{ASCA} data. The author found that the subsolar metallicity found in previous studies is a fitting artifact caused by assuming an isothermal model. This issue was referred to as the 'iron bias' and was rectified by fitting with a more appropriate mode, namely, the 2T model. The mean metallicity of the sample of \cite{Buote_2000} became a factor $2.6$ larger than the value obtained by using the 1T model. On the other hand, the central abundance drop in the Centaurus cluster was previously observed in other studies \citep[e.g.,][]{SandersFabian_2002, Matsushita_2007, Sanders_2016, Lakhchaura_2019, Liu_2019} and was linked to the dust depletion scenario, such that a significant fraction of metals in the hot phase cools down and then get incorporated into the dust grains from the central BCG \citep{Panagoulia_2015, Mernier_2017}. Afterward, the metal-rich dust grains are displaced to larger radii by the AGN feedback \citep{Simionescu_2009, Kirkpatrick_2011}. While we confirm the iron bias affecting this first bin, however, due to the limitation from the point spread function (PSF) and statistics, the analyzed bin is larger than previous studies \citep[e.g.,][]{Fukushima_2022}, which may not be ideal for fully resolving the central metallicity drop \citep{Gatuzz_2023}.

\subsubsection{Eastern and western edges}
In the 4-pixel GGM-filtered image, we observed the eastern and western edges that are $\sim2.1'$ and $3.6'$ from the cluster center. These edges are also evident from the surface brightness profiles (Fig.~\ref{fig:SBprofiles_infull} for $0.2-2.3~\mathrm{keV}$ or \ref{fig:SBprofiles_inhi} for $1.0-2.3~\mathrm{keV}$). For instance, in the eastern profile of the $0.2-2.3~\mathrm{keV}$ band, the significance surface brightness with respect to the full azimuthal value at $1.5'$ (before the edge) is $5.7\sigma$, then the following bin (centered at $2.0'$), the significance drops to $-2.1\sigma$. For the western edge, we noticed a $3.8\sigma$ excess at $2-3'$. At the following bins centered at $3.5$ and $4.0'$, the significance values drop to $1.3$ and $-2.3\sigma$, respectively. \cite{Sanders_2016} describe these features as cold fronts \citep{Markevitch_2007}, as they are accompanied by temperature jumps and surface density drops in front of the edges \citep[also see][]{SandersFabian_2002, Fabian_2005}. These cold fronts, as well as the east-west asymmetry appearance and metal distributions in the core, are common signatures of sloshing motions that could be triggered by the interaction between Cen 30 and Cen 45 \citep{Fabian_2005, Sanders_2016}. Other sloshing examples seen in observations include the Perseus cluster \citep{Simionescu_2012, Urban_2014}, Virgo cluster \citep{Gatuzz_2022Virgo}, A2657 \citep{Botteon_2024}, A2029 \citep{Paterno_2013}, A2142 \citep{Markevitch_2000, Liu_2018}.

\subsubsection{Western filamentary excess}
Another feature is the filament-like excess in the west of the cluster center. The feature is only apparent in the residual image (Fig.~\ref{fig:residual}). Note that the eRASS:5 Centaurus residual image is strikingly similar to the \rosat PSPC relative excess image \citep[right panel of Fig.~3 in ][]{Churazov_1999} and the \xmm residual image \citep[Fig.~7 in][]{Walker_2013}, which were only feasible up to $r<R_{2500}$. This filament-like excess seems to be related to the S0 galaxy NGC 4696B ($z=0.0104$), which is located in the west, near the boundary of the $R_{2500}$ (yellow circle in Fig.~\ref{fig:residual}). The NGC 4696B galaxy is the third brightest galaxy after NGC 4696 and NGC 4709 in the optical band. As observed in the residual image, the X-ray position of NGC 4696B is slightly offset to its optical position toward the northeast. This excess emission structure was speculated to be ram pressure-stripped gas from NGC 4696B \citep{Churazov_1999}. Higher metallicity in the innermost bin in the north (by a factor of $1.6$ and a significance of $2.7\sigma$ to the south) may also be contributed by the stripped-gas \citep{Walker_2013}.

\subsubsection{Eastern excess}\label{sec:res_Eexcess}
The residual image (Fig.~\ref{fig:residual}) reveals a bow-shaped excess in the east of NGC 4696, which is also evident in the \rosat PSPC relative excess emission \citep{Churazov_1999} and the \xmm residual image \citep{Walker_2013}. Completing the detection view by previous instruments, we show that the eastern excess extends beyond the $R_{2500}$ boundary and diminishes as it approaches $R_{500}$. Up to $R_{2500}$, we acquired an average surface brightness excess with $4.1\sigma$ significance in the eastern sector with respect to the average surface brightness values. Between $R_{2500}$ and $0.6R_{500}$, the average eastern significance drops to $2.5\sigma$. The excess beyond $R_{2500}$ in the east is confirmed as the eastern extension in the 32-pixel GGM-filtered image (Fig.~\ref{fig:GGM}, right), which also coincides with the location of a galaxy overdensity patch of the Cen 45 member galaxies (Fig.~\ref{fig:galnumdensb}).
\par
From the spectral analysis, excess emission is also observed. We found that the normalizations per area in the east are higher than the full annuli and other sectors, except for the outermost bin ($R_{500}-R_{200}$). With respect to the full azimuthal values, there is a $31\%$ and $36\%$ enhancement of $5.4\sigma$ at the bin before ($10.0-15.0'$) and after the center of Cen 45 ($15-20.0'$), respectively. The excess peaks at $20.0-25.4'$ bin with $45\%$ and $7.7\sigma$ and continues until the $0.7-1.0R_{500}$ bin with a relative difference of $36\%$ and $2.6\sigma$ significance, showing that the Cen 45 substructure extends even further out than previously measured. \cite{Gatuzz_2022} found a higher ICM redshift with respect to the Cen 30 optical redshift at a $r>15'$ bin, east of Cen 45, which signifies that the eastern extension may be a tail of Cen 45 emerging from the merger activity.
\par
At the $15-20.0'$ bin, we notice that the temperatures in the east ($3.18_{-0.29}^{+0.41}~\mathrm{keV}$) and south ($3.41_{-0.51}^{+0.98}~\mathrm{keV}$) are $0.8~\mathrm{keV}$ and $1.0~\mathrm{keV}$ higher to their counterpart bins in the west and north, with a significance of $2.3\sigma$ for the east to west and $1.6\sigma$ for the south to north comparisons. The temperature values in these bins are consistent with previous temperature enhancement findings in the same directions by other instruments. Using \textit{ASCA} observations, \cite{Churazov_1999} and \cite{Furusho_2001} found a temperature value of $4.4\pm0.2~\mathrm{keV}$ and $4.3\pm0.2~\mathrm{keV}$, respectively, while \cite{Walker_2013} reported a hot component of $5.0\pm0.4~\mathrm{keV}$ from their analysis using \xmm observations. Using the eROSITA-\xmm temperature relation of \cite{Migkas_2024}, we converted the eastern and southern temperatures at the $15-20.0'$ bin. The calculated values (without scatters) are in agreement within $<1.6\sigma$ with the reported \xmm temperature by \cite{Walker_2013}. The temperature enhancement in this location is consistent with shock-heated gas caused by the interaction between Cen 30 and Cen 45.

\subsubsection{$R_{2500}<r<R_{500}$}

The spectral parameter profiles of the different sectors decrease outward ($R_{2500}$) with roughly the same shape but with some fluctuations across the different sectors. The most noticeable is the previously mentioned excess normalization per area in the east (see Sect.~\ref{sec:res_Eexcess}). Other fluctuation is observed at the southern $0.7-1.0R_{500}$ ($42.2-59.0'$) bin, where the temperature is higher ($2.28_{-0.48}^{+0.91}~\mathrm{keV}$) in comparison to its northern counterpart bin ($1.59_{-0.11}^{+0.16}~\mathrm{keV}$), with a $1.4\sigma$ significance. The average best-fit temperature in the $R_{2500}-0.7R_{500}$ ($25.4-42.2'$) is $2.2~\mathrm{keV}$, with a range of $1.9-2.4~\mathrm{keV}$. For the $0.7-1.0R_{500}$ ($42.2-59.0'$) bin, the best-fit temperature average is $1.8~\mathrm{keV}$, ranging from $1.1$ to $2.5~\mathrm{keV}$.
\par
In the temperature profile panels of Fig.~\ref{fig:specprofiles}, we compared the outskirts temperature results with the expected temperature profiles, i.e., temperature profile fitted from \suzaku measurements \citep{Reiprich_2013} and the average galaxy cluster temperature profile from $N$-body plus hydrodynamics cosmological simulations \citep{Burns_2010}. The temperatures of the Centaurus cluster in $R_{2500}<r<R_{500}$ appear to drop smoothly and the values are in good agreement with the profiles from simulation and fit of \suzaku outskirts measurements.
\par
The metallicity values across the sectors in the $R_{2500}-0.7R_{500}$ ($25.4-42.2'$) are in very good agreement with each other with an average value of $0.21Z_\odot$. For the $0.7-1R_{500}$ ($42.2-59.0'$) bin, the metallicity values in the north, south, and west are consistent with each other, with their values lying between $0.07Z_\odot$ and $0.40Z_\odot$. Meanwhile, we found lower metallicity in the east ($0.03_{-0.03}^{+0.04}Z_\odot$) with a $2.3\sigma$ significance deviation with respect to the western sector at the same radial bin. \cite{Walker_2013Suzaku} analyzed \suzaku observations of Centaurus cluster along the north-western strip out to $0.95R_{200}$. At $45-50'$, they report a metallicity of $0.1Z_\odot$, which is broadly in agreement with the value reported in this work. In general, our metallicity values in this regime are in good agreement with other galaxy cluster studies \citep[e.g.,][]{Akamatsu_2011, Bulbul_2016, Urban_2017, Ghizzardi_2021}, that report a metallicity value of $\sim\!0.3Z_\odot$ at a similar distance range.

\subsubsection{$r>R_{500}$}
In the $R_{500}<r<R_{200}$ regime, we measured cluster surface brightness above the CXB level with significance values of the full azimuthal direction spreading between 3.0 and $7,2\sigma$ (see bottom plot of Fig.~\ref{fig:SBprofiles_outhi} or \ref{fig:SBprofiles_outfull}). At $r> R_{200}$, we observed various levels of eROSITA Bubbles contamination in the different sectors from the $0.2-2.3~\mathrm{keV}$ surface brightness profiles (Fig.~\ref{fig:SBprofiles_outfull}), where the strongest is in the east, milder in the north and south, and none in the west. The eROSITA Bubble contamination diminishes in the $1.0-2.3~\mathrm{keV}$ profiles (Fig.~\ref{fig:SBprofiles_outhi}). In this higher energy band profile, we detected a $3.5\sigma$ cluster emission above the CXB at $90'~(\sim\!R_{200})$, and then followed by $2.9\sigma$ at $100'~(\sim\!1.1R_{200})$.
\par
In this regime, the spectral analysis was performed out to $R_{200}$. There is a temperature drop by a factor of $\sim\!3$ from the peak temperature to this outermost bin. The temperature values in the different sectors at $R_{500}-R_{200}$ are consistent with each other, with an average of $1.2~\mathrm{keV}$. The outskirts temperatures closely adhere to the predicted temperature profiles derived from simulations and fit from \suzaku outskirts measurements. The absence of temperature deviations, such as excess, in the outskirts of the Centaurus cluster corresponds to the lack of large-scale filaments connected to the cluster \citep["connectivity", e.g.,][]{Gouin_2021}, which is in agreement with the results of image inspections and optical redshift analysis. As reported in other works \citep[e.g.,][]{Kawaharada_2010, Veronica_2024}, connectivity with large-scale structures may induce faster thermalization along the connected direction of the cluster outskirts.
\par
The metallicity values acquired in the northern and southern sectors are slightly lower than in the eastern and western sectors. The values range between $0.003$ and $0.08Z_\odot$ for the former and between $0.07$ and $0.21Z_\odot$ for the latter. Nevertheless, our findings in this regime are consistent within the measurement ranges of other galaxy clusters; for example \suzaku outskirts metallicity of A2204 \citep{Reiprich_2009}, A1413 \citep{Hoshino_2010}, Perseus cluster \citep{Simionescu_2011, Werner_2013}, and ten nearby clusters \citep{Urban_2017}, as well as the \xmm measurement of the northern outskirts of the Virgo cluster \citep{Urban_2011}.

\section{Summary and conclusions}\label{sec:conclude}
Utilizing the combined five eROSITA All-Sky Survey data (eRASS:5), we performed comprehensive imaging and spectral analyses out to $2R_{200}~(\sim\!181')$ and $R_{200}~(\sim\!91')$, respectively, unveiling the first view of Centaurus sampling the whole azimuth beyond $30'$. Our data reduction and image correction steps include flare filtering, Galactic absorption correction, and exposure correction accounting for the different types of eROSITA's telescope modules (see Sect.~\ref{sec:imag_analysis}). The steps were realized in different energy bands to assess the eROSITA Bubble emission contamination in the field. We applied various image manipulation methods employing the final corrected image products, such as wavelet filtering, adaptive smoothing, and GGM filtering to search for cluster features. We quantified the features by calculating and comparing surface brightness profiles of different energy bands ($0.2-2.3~\mathrm{keV}$ and $1.0-2.3~\mathrm{keV}$) in full azimuth and four sectors based on the morphology of the cluster. We conducted rigorous sky background spectral analysis to constrain local sky fore- and background components. We developed a technique to spatially subtract the sky background across the FoV from the image. Last but not least, from spectral analysis, we acquired the ICM parameter profiles, namely, normalizations, temperature, and metallicity in various directions. We compared our inner region results ($r<R_{2500}$) with previous studies. In the outskirts ($r>0.3R_{200}$), we compared the eRASS:5 temperature profile with other cluster measurements and simulations. Additionally, we compiled a galaxy redshift catalog from a public database and carried out galaxy member analysis. We correlated the cluster galaxy member distribution with X-ray emission. We summarized our results below:

\begin{itemize}
    \item We observed the known bimodality in the distribution of the redshifts of the galaxies within $R_{200}$ of the Centaurus cluster. The peaks of the bimodality are at $z=0.0104$ that corresponds to the Cen 30 substructure at the cluster center, and at $z=0.0149$ for the Cen 45 substructure. The peak of galaxy member number density of each substructure is centered at their respective brightest galaxy; NGC 4696 for Cen 30 and NGC 4709 for Cen 45 ($14.9'$ east of the cluster center).
    \item eRASS:5 images, as well as surface brightness and spectral analyses reveal the previously known features in the cluster inner region ($r<R_{2500}\approx25.4'$), for instance, the soft filament emission originated from the nucleus (plume), temperatures and metallicities asymmetry in the core, and the eastern and western cold fronts. These features are associated with AGN feedback from NGC 4696 galaxy and the sloshing motion triggered by substructure merger. Additionally, we also observed the western filamentary excess that seems to be the ram-pressure-stripped gas from NGC 4696B (the third optically brightest galaxy in the Centaurus cluster). This feature might be responsible for the higher core metallicity in the north.
    \item We found that the spectra of the innermost bin of $0.0-0.8'$ ($r<10.7~\mathrm{kpc}$) is described better with a multi-temperature model (2T) than an isothermal model (1T). The core temperature drops from $1.4~\mathrm{keV}$ to $1.0~\mathrm{keV}$. Meanwhile, the metallicity value increases from $0.67_{-0.05}^{+0.05}Z_\odot$ to $1.59_{-0.18}^{+0.22}Z_\odot$, implying that this bin suffers from an iron bias. 
    \item We traced the X-ray eastern excess emission up to $R_{500}$, revealing the full extension of the Cen 45 substructure. The eastern enhancement (normalization per area) peaks at the $20.0-25.4'$ bin ($\sim\!5.1-10.5'$ behind the Cen 45 center) with a relative difference of $45\%$ and significance of $7.7\sigma$ to the full annulus value, then continues until the $42.2-59.0'$ ($0.7-1.0R_{500}$) with a $36\%$ and $2.6\sigma$. The excess behind Cen 45 might be the ram-pressure-striped gas from the substructure. Moreover, we confirmed the previously detected temperature enhancements in the east and south, which are consistent with shock-heated gas resulting from the interaction between Cen 30 and Cen 45.
    \item We measured significant emission beyond $R_{200}$, that is surface brightness value of $3.5\sigma$ above the CXB level at $R_{200}$ and $2.9\sigma$ at $\sim\!1.1R_{200}$.
    \item The cluster temperature profiles in different sectors show a similar trend, namely a drop in the core and a peak in the intermediate radial range. A drop by a factor $\sim\!3$ from the peaks to the outermost bin ($R_{500}-R_{200}$) is seen in various sectors. The measured temperature profiles in the outskirts ($r>0.3R_{200}$) follow closely the profiles from simulations and fit from \suzaku cluster outskirts measurements.
    \item The metallicity peaks in the center and drops rapidly outwards. In the outskirts ($R_{500}-R_{200}$), a little asymmetry is observed between the east-west (semi-major axis) and north-south (semi-minor axis) measurements, where the first is higher than the latter. The range of outskirts metallicity in the eastern and western sectors is between $0.07$ and $0.21Z_\odot$, while in the northern and southern sectors is between $0.003$ and $0.08Z_\odot$. Higher metallicity in the east-west direction might be related to the merger direction. We report that the metallicity estimates of these sectors are in good agreement with the other cluster outskirts measurements.
\end{itemize}

\begin{acknowledgements}
      We thank the anonymous referee for their valuable feedback that helped improve the manuscript.
      Funded by the Deutsche Forschungsgemeinschaft (DFG, German Research Foundation) – 450861021.
      A.V. is a member of the Max-Planck International School for Astronomy and Astrophysics (IMPRS) and of the Bonn-Cologne Graduate School for Physics and Astronomy (BCGS), and thanks for their support.
      T.R. acknowledges support from the German Federal Ministry of Economics and Technology (BMWi) provided through the German Space Agency (DLR) under project 50 OR 2112.
      M.Y. acknowledges support from the Deutsche Forschungsgemeinschaft through the grant FR 1691/2-1.
      This work was supported in part by JSPS KAKENHI Grant Number 20K04027 (NO).
      This work is based on data from eROSITA, the soft X-ray instrument aboard SRG, a joint Russian-German science mission supported by the Russian Space Agency (Roskosmos), in the interests of the Russian Academy of Sciences represented by its Space Research Institute (IKI), and the Deutsches Zentrum für Luft- und Raumfahrt (DLR). The SRG spacecraft was built by Lavochkin Association (NPOL) and its subcontractors, and is operated by NPOL with support from the Max Planck Institute for Extraterrestrial Physics (MPE). The development and construction of the eROSITA X-ray instrument was led by MPE, with contributions from the Dr. Karl Remeis Observatory Bamberg and ECAP (FAU Erlangen-Nuernberg), the University of Hamburg Observatory, the Leibniz Institute for Astrophysics Potsdam (AIP), and the Institute for Astronomy and Astrophysics of the University of Tübingen, with the support of DLR and the Max Planck Society. The Argelander Institute for Astronomy of the University of Bonn and the Ludwig Maximilians Universität Munich also participated in the science preparation for eROSITA. The eROSITA data shown here were processed using the eSASS software system developed by the German eROSITA consortium.
      This research has made use of the NASA/IPAC Extragalactic Database (NED), which is funded by the National Aeronautics and Space Administration and operated by the California Institute of Technology.
\end{acknowledgements}

%
%
\bibliographystyle{aa}
\bibliography{list_bib}

\begin{appendix}
\section{$N_\mathrm{H}$ map}\label{App:A}

\begin{figure}[!h]
\centering
\includegraphics[width=\columnwidth, trim=0cm 0cm 0.05cm 0cm,clip]{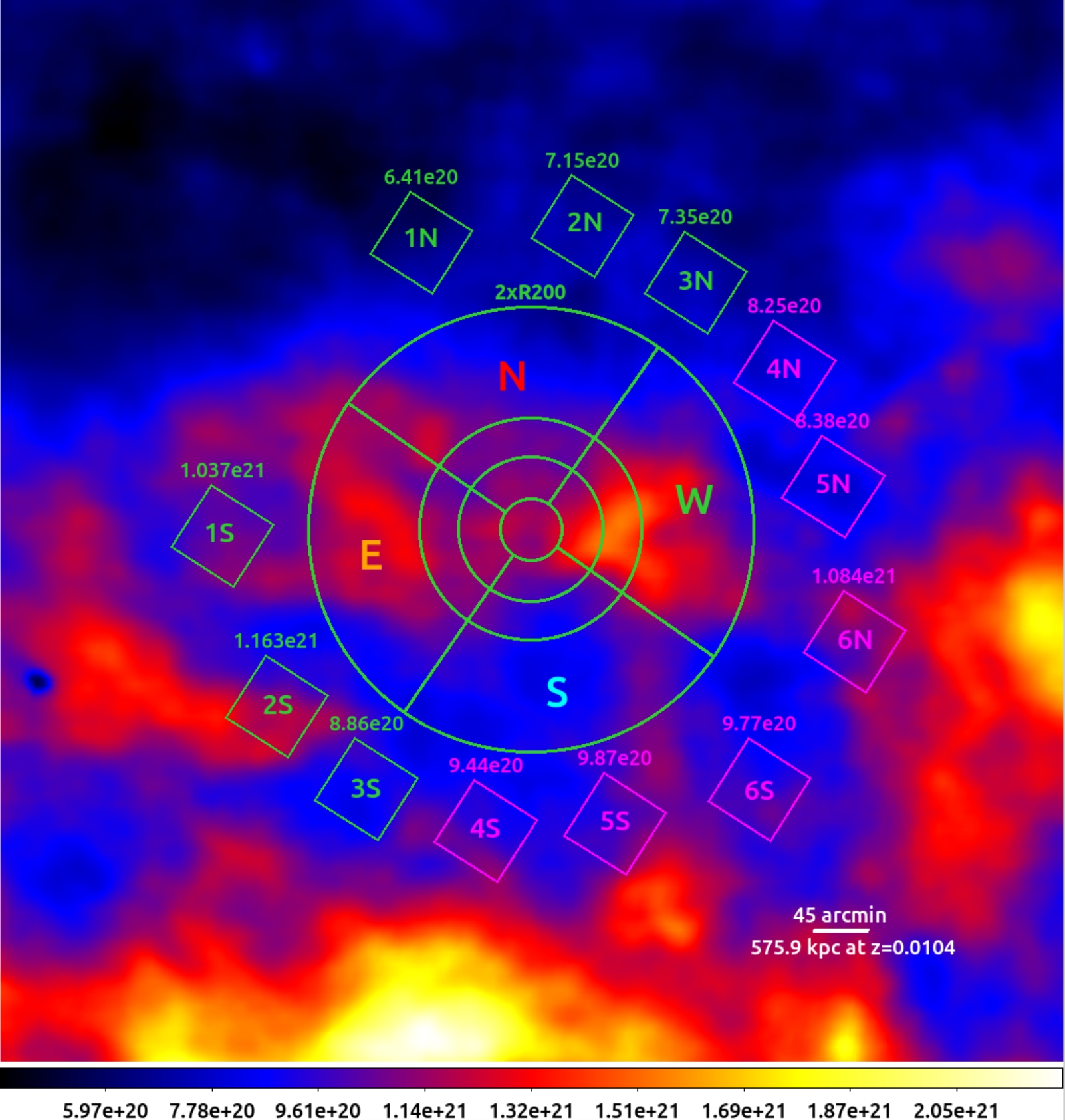}
\hspace{5pt}
\caption{The total $N_\mathrm{H}$ map (combination of HI4PI all-sky Galactic neutral Hydrogen map \citealp{HI4PI_2016} and $N_\mathrm{H2}$ map from \citealp{Willingale_2013}) of the Centaurus cluster field. The configurations of the science and sky background analyses are overlaid on top (see Sect.~\ref{sec:drsteps}).}
\label{fig:nh}
\end{figure}

\section{X-ray Images}
\begin{figure*}[!h]
\centering
\includegraphics[width=0.85\textwidth,trim=0.25cm 0.0cm 0.04cm 0.0cm,clip]{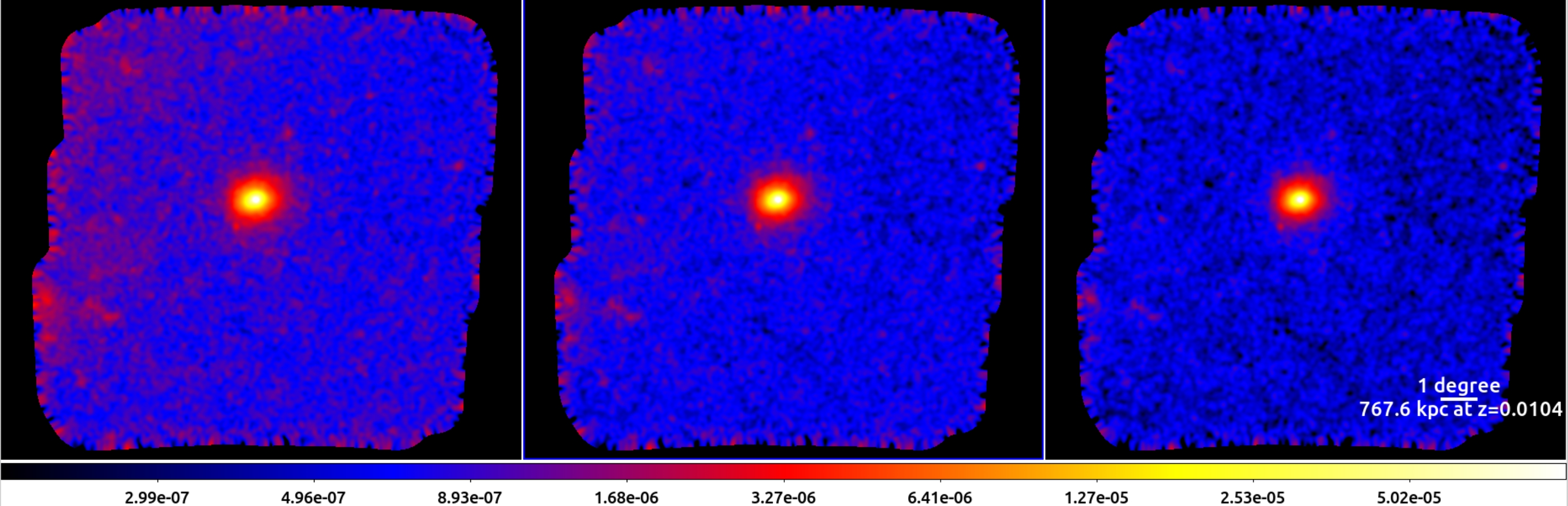}
\caption{Same as Fig.~\ref{fig:conf}, but in different energy bands. \textit{Left:} $0.8-2.3~\mathrm{keV}$ band. \textit{Middle:} $0.9-2.3~\mathrm{keV}$. \textit{Right:} $1.0-2.3~\mathrm{keV}$.}
\label{fig:erobub}
\end{figure*}

The eRASS:5 fully-corrected images in the higher energy bands are shown in Fig.~\ref{fig:erobub}. The images have been adaptively smoothed with S/N set to 45.

\subsection{CXB map and CXB-subtracted image}\label{app:erobub}
\begin{figure*}[h!]
\centering
\includegraphics[width=0.57\textwidth, trim=0cm 0cm 0.0035cm 0cm,clip]{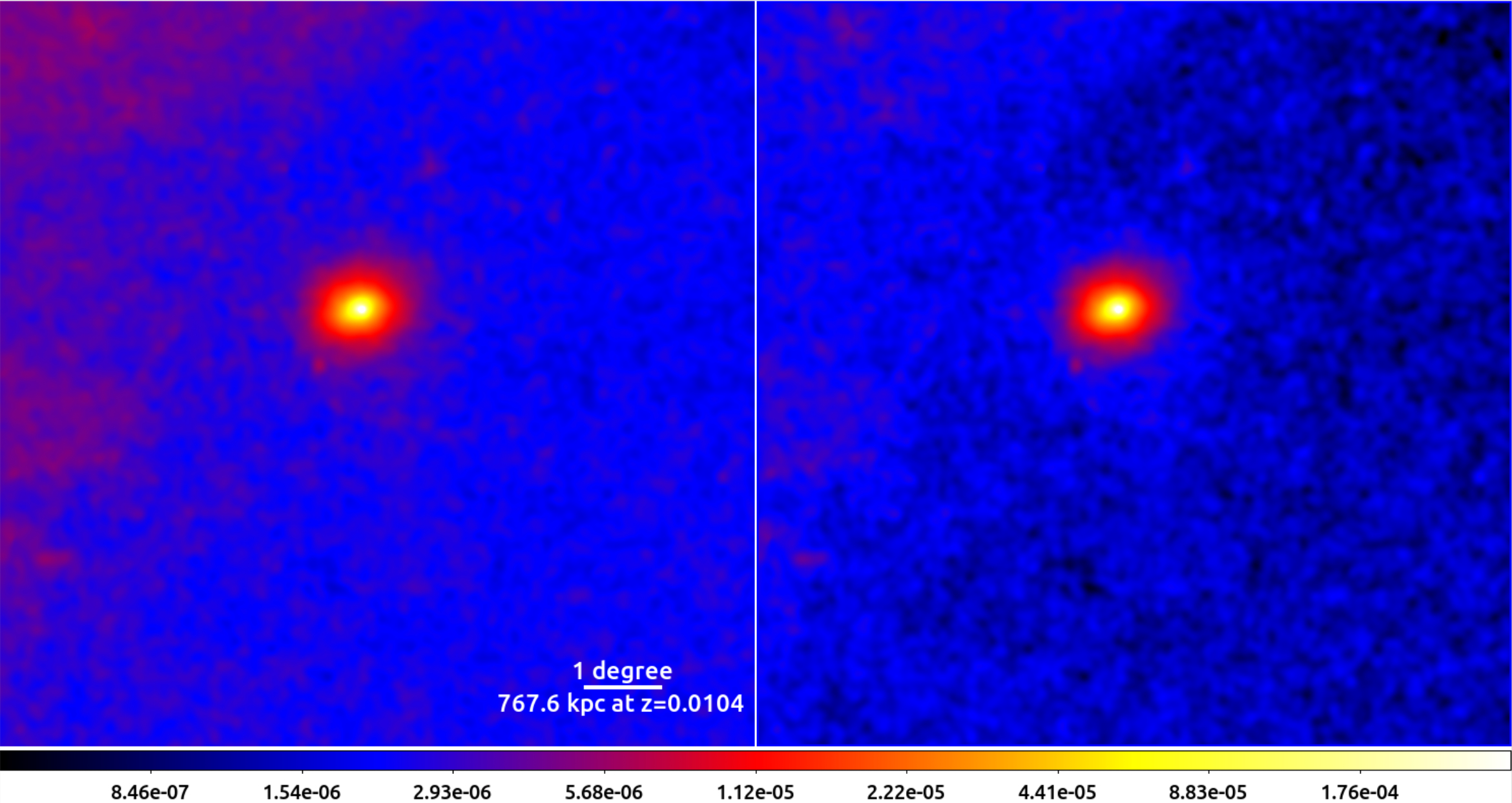}
\includegraphics[width=0.285\textwidth, trim=0.1cm 0cm 0.05cm 0.1cm,clip]{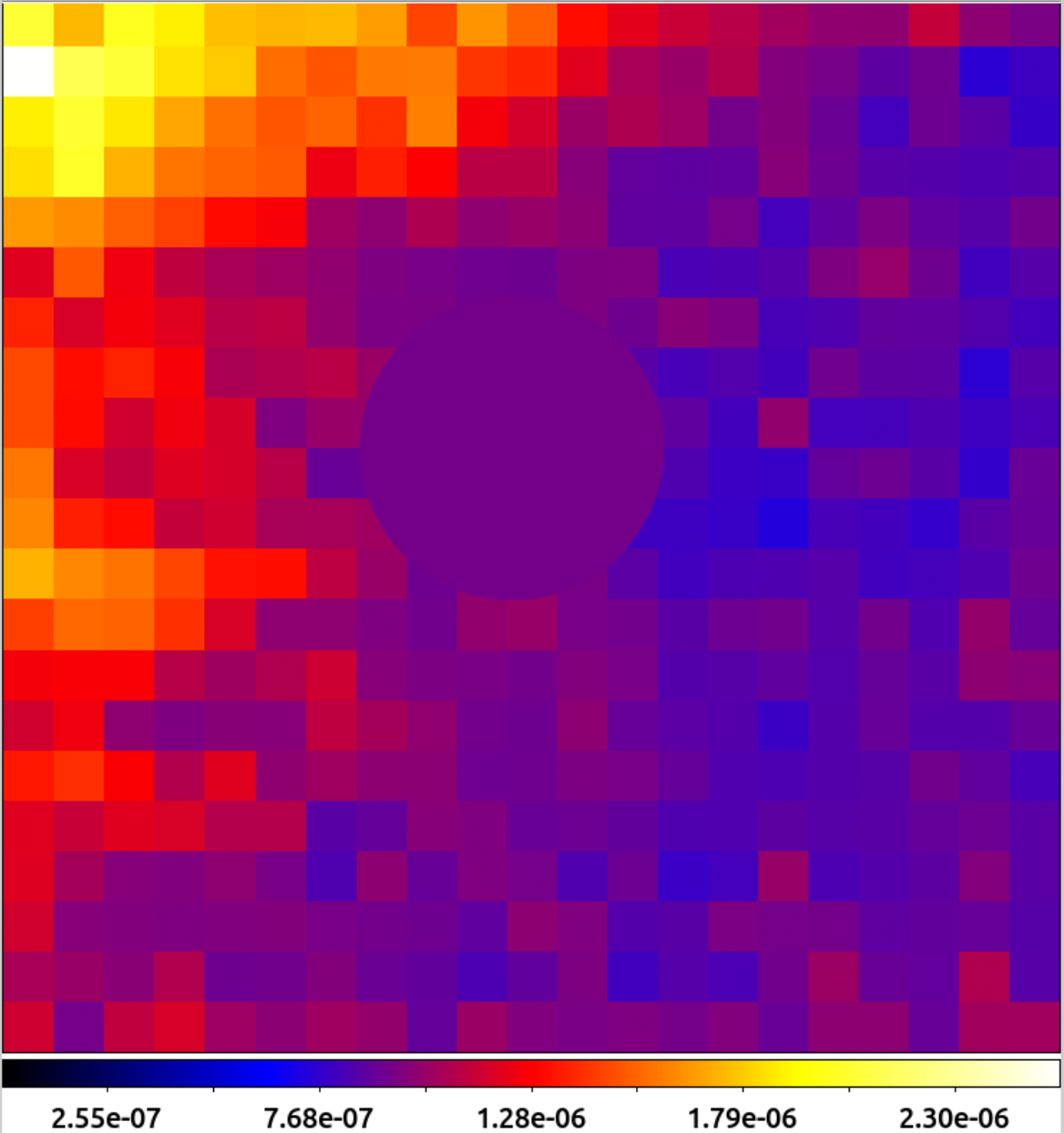}
\caption{CXB subtraction attempt in the Centaurus cluster field. \textit{Left:} Same as Fig.~\ref{fig:conf}. \textit{Middle:} CXB-subtracted, fully-corrected image (\textit{left} minus \textit{right}). The image was then adaptively smoothed with the same S/N as the left image. \textit{Right:} CXB map including the eROSITA Bubble emission in the $0.2-2.3~\mathrm{keV}$. Note that the color bar range of the right panel is different than the other panels.}
\label{fig:erobub_Sub}
\end{figure*}

In an attempt to spatially remove the sky background emission from the image, particularly the eROSITA Bubble emission, we performed spectral fitting of 409 sky background boxes located $r>R_{200}$. Each box has a size of $0.5^\circ\times0.5^\circ$. During the fit, we omitted the cluster emission. All sky fore- and background components were frozen to the values listed in Table~\ref{tab:sky_BG}, except for the eROSITA Bubble normalization that shows significant variation across the FoV. From each fitting, we acquired the count-rates value of the on-chip TMs in the energy band of $0.2-2.3~\mathrm{keV}$. We then computed the CXB count rates per pixel by taking the average on-chip TM count-rate value normalized by the corresponding area of the box. The resulting CXB map is displayed in the right panel of Fig.~\ref{fig:erobub_Sub}. For the $r<R_{200}$, we computed the average of the values from the immediate surrounding boxes. The CXB-subtracted, fully-corrected image in the $0.2-2.3~\mathrm{keV}$ is shown in the middle panel of Fig.~$\ref{fig:erobub_Sub}$. While the fore-/background emission is drastically reduced, making the cluster outskirts more visible, a faint residual from the eROSITA Bubble can be seen in the left part of the image. The residual is likely the hotter phase of the structure, as can be concluded from the yellowish emission (red-soft and green-intermediate channels) in the RGB image (Fig.~\ref{fig:rgb}) and apparent emission at higher energy cut image (left panel of Fig.~\ref{fig:erobub}). However, we could not constrain this hotter component from the spectra as described in Sect.~\ref{sec:erobubble}. Nevertheless, we present the result of the first attempt and method of spatially removing the CXB contribution from the image, which can be further improved for future large-scale environmental studies.

\section{X-ray surface brightness profiles}
\begin{figure*}
\centering
\subfloat[$0 \leq r \leq R_{500}$]{\includegraphics[width=0.4\textwidth,trim=0.1cm 0.2cm 0.5cm 0.3cm,clip]{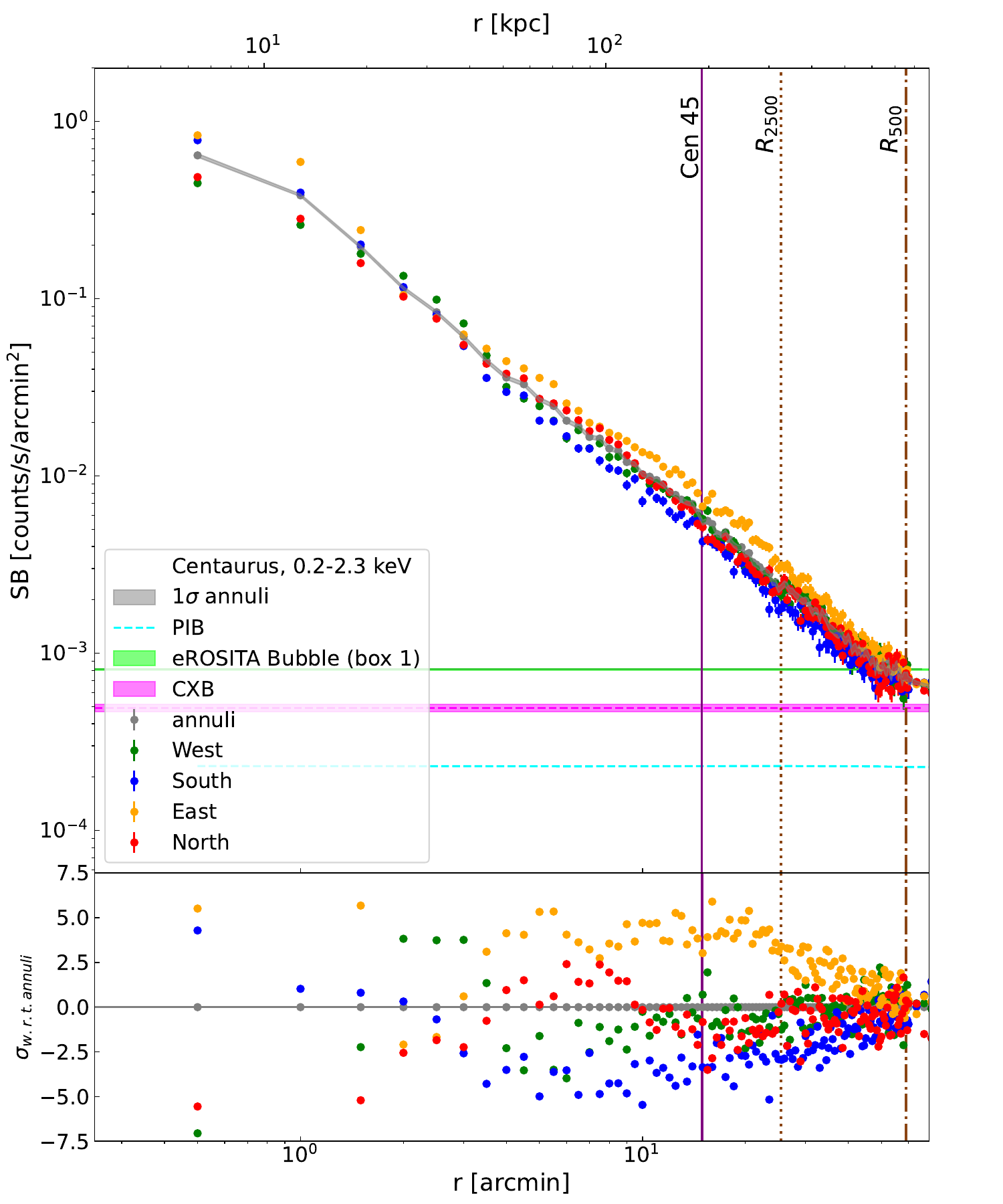}\label{fig:SBprofiles_infull}}
\subfloat[$R_{500}\leq r \leq 2R_{200}$]{\includegraphics[width=0.4\textwidth,trim=0.1cm 0.25cm 0.5cm 0.3cm,clip]{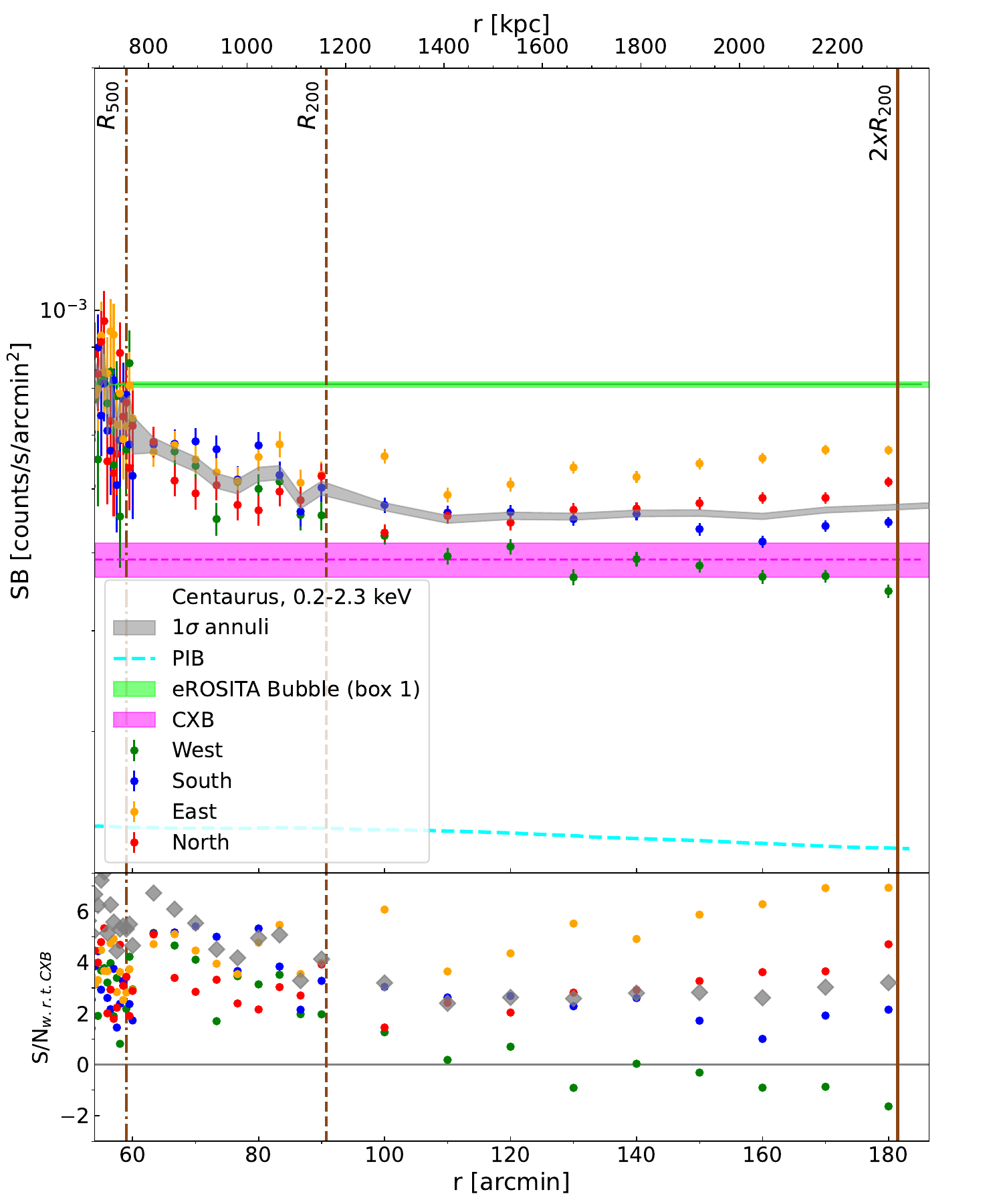}\label{fig:SBprofiles_outfull}}
\caption{Same as Fig.~\ref{fig:SBprofiles_hi}, but in the $0.2-2.3~\mathrm{keV}$ band.}
\label{fig:SBprofiles}
\end{figure*}

The X-ray surface brightness profiles in the $0.2-2.3~\mathrm{keV}$ band are shown in Fig.~\ref{fig:SBprofiles}.

\section{Spectral Analysis}\label{App:spectro}
\begin{figure*}[h!]
\centering
\includegraphics[width=\textwidth,trim=4.55cm 0.5cm 4.55cm 1.0cm,clip]{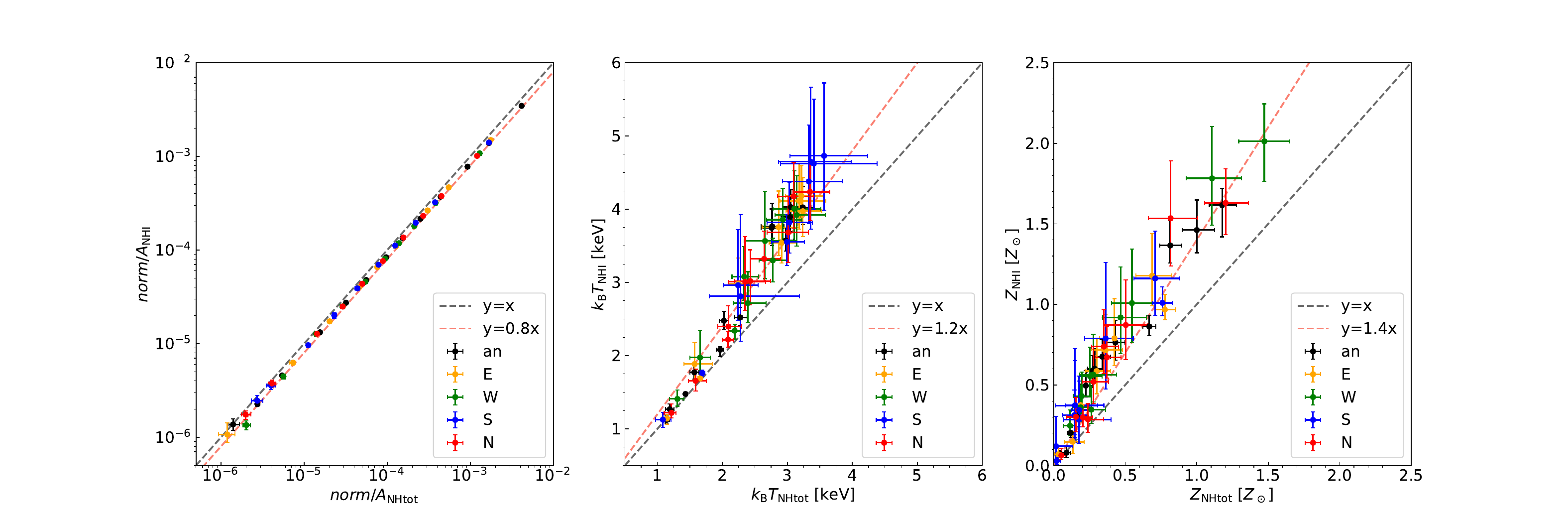}
\caption{Comparison plots of the different spectral parameters of the Centaurus cluster constrained from eRASS:5 data. The data were fitted using different absorption values, as discussed in Sect.~\ref{sec:spectro_results}. The y-axes represent values obtained using the neutral hydrogen column density ($N_\mathrm{HI}$), while the x-axes from the total hydrogen column density ($N_\mathrm{Htot}$). Gray dashed lines indicate the 1:1 ratio, and the orange dashed lines represent offset lines, as detailed in the legends.}
\label{fig:nhI-nhtot}
\end{figure*}

\begin{figure*}[h!]
\centering
\includegraphics[width=0.41\textwidth,trim=0.5cm 0.5cm 1.5cm 0.5cm,clip]{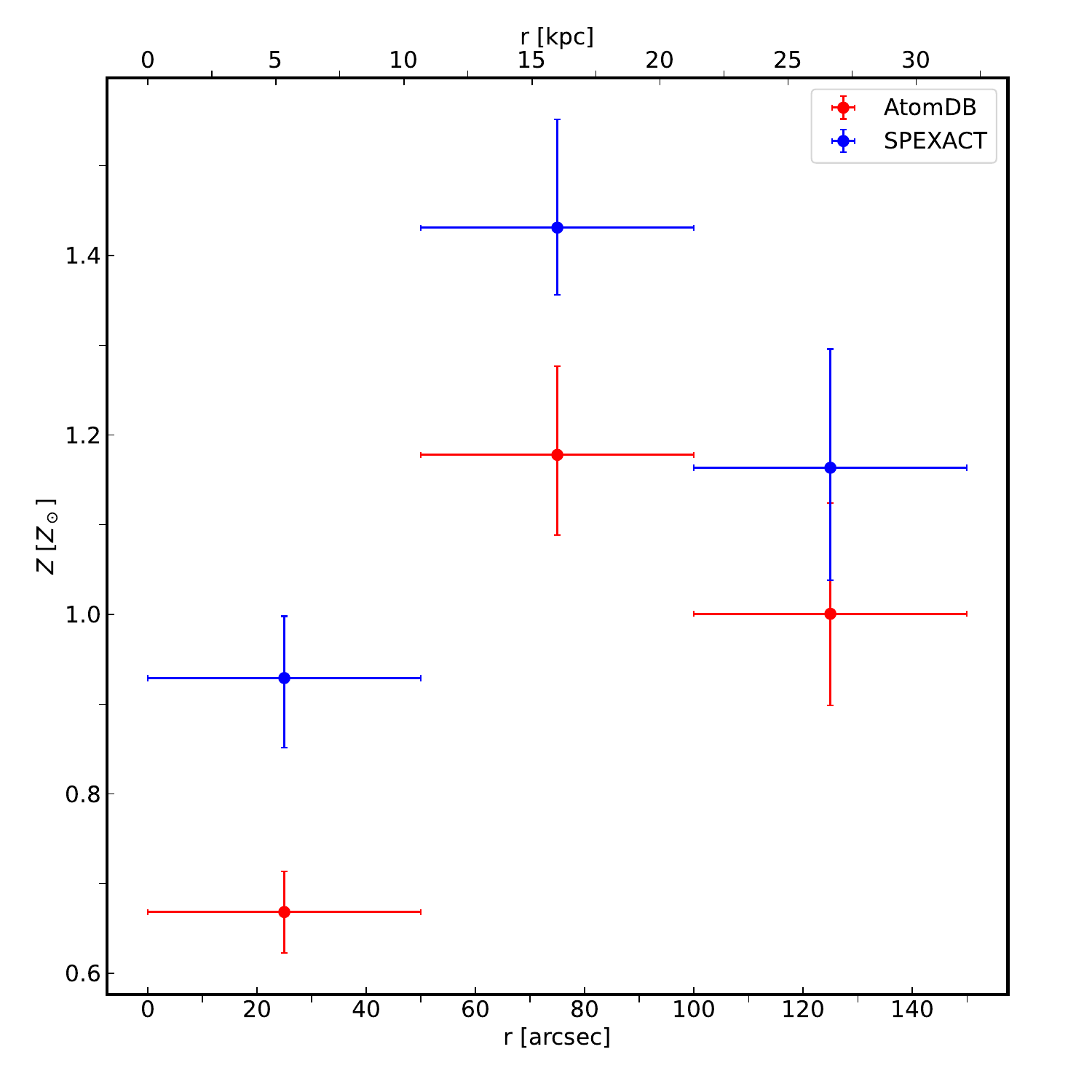}
\includegraphics[width=0.41\textwidth,trim=0.5cm 0.5cm 1.5cm 0.5cm,clip]{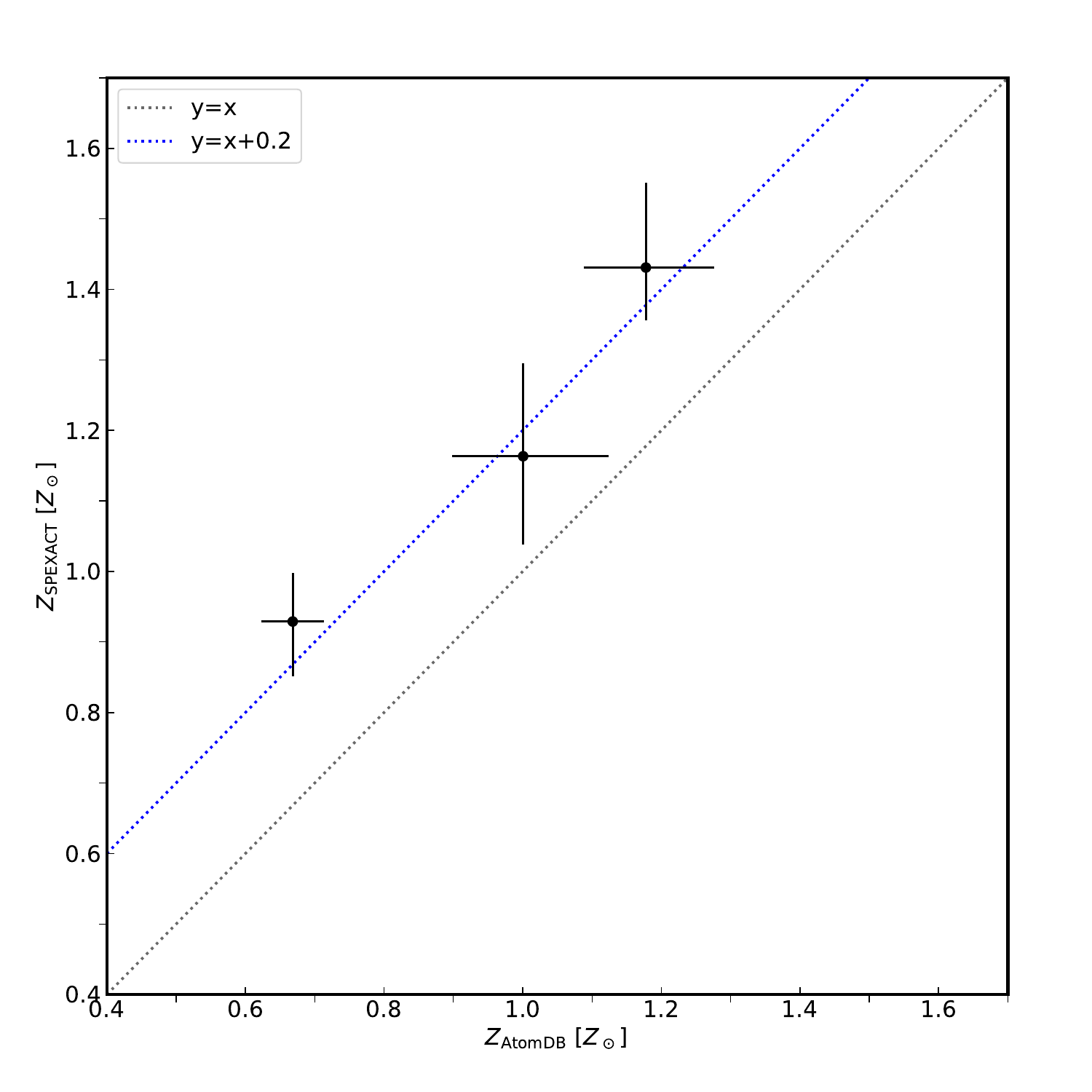}
\caption{\textit{Left:} Centaurus core metallicity profiles using AtomDB version 3.0.9 (red) and SPEXACT version 3.05.00 (blue). \textit{Right:} Comparison plot of the metallicities using the two atomic databases. The gray dotted line is the 1:1 line, while the blue is shifted by 0.2.}
\label{fig:spex}
\end{figure*}

We present the results of the single-temperature spectral analysis (1T) in Table~\ref{tab:spectro} and the results of the two-temperature spectral analysis (2T) in Table~\ref{tab:spectro2T}. Additionally, comparison plots showing $N_\mathrm{HI}$ against $N_\mathrm{Htot}$ fits, as well as AtomDB against SPEXACT fits, are displayed in Fig.~\ref{fig:nhI-nhtot} and Fig.~\ref{fig:spex}, respectively.
\begin{table*}
    \centering
    \caption{The eRASS:5 spectral analysis results (1T fit) of the Centaurus cluster in full azimuthal annuli and various sectors out to $R_{200}$.}
    \resizebox{\textwidth}{!}{\begin{tabular}{c c c c c c || c c c c c c}
    \hline
    \hline
\multirow{2}{*}{Sector} & bin & $norm$ & $k_\mathrm{B}T$ & $Z$ & \multirow{2}{*}{stat/dof} & \multirow{2}{*}{Sector} & bin & $norm$ & $k_\mathrm{B}T$ & $Z$ & \multirow{2}{*}{stat/dof}\\
& $[\mathrm{arcmin}]$ & $[10^{-5}$ cm$^{-5}/$arcmin$^2]$ & $[$keV$]$ & $[Z_{\odot}]$ & &  & $[\mathrm{arcmin}]$ & $[10^{-5}$ cm$^{-5}/$arcmin$^2]$ & $[$keV$]$ & $[Z_{\odot}]$ & \\
\hline \\[-1.7ex]
\multirow{12}{*}{Annuli} & $0.0-0.8$ & $414.69_{-13.79}^{+15.27}$ & $1.43_{-0.01}^{+0.03}$ & $0.67_{-0.05}^{+0.05}$ & 7384.2/7294 & \multirow{12}{*}{Annuli} & \multirow{3}{*}{$0.0-2.5$} & \multirow{2}{*}{$162.77_{-3.07}^{+2.99}$} & \multirow{2}{*}{$1.78_{-0.02}^{+0.02}$} & \multirow{2}{*}{$0.84_{-0.04}^{+0.04}$} & \multirow{2}{*}{8248.9/7974} \\[5pt]
& $0.8-1.7$ & $167.55_{-5.55}^{+5.48}$ & $1.97_{-0.06}^{+0.05}$ & $1.18_{-0.09}^{+0.10}$ & 7558.4/7498& & & & & &\\[5pt]
& $1.7-2.5$ & $92.66_{-3.22}^{+3.10}$ & $2.28_{-0.08}^{+0.11}$ & $1.0_{-0.10}^{+0.12}$ & 7443.1/7433& & & & & &\\[5pt]
& $2.5-5.0$ & $43.98_{-0.9}^{+0.9}$ & $2.98_{-0.11}^{+0.12}$ & $0.81_{-0.07}^{+0.08}$ & 8072.7/8066& & & & & &\\[5pt]
& $5.0-7.5$ & $25.14_{-0.51}^{+0.5}$ & $3.04_{-0.14}^{+0.18}$ & $0.34_{-0.05}^{+0.06}$ & 8143.5/8127& & & & & &\\[5pt]
& $7.5-10.0$ & $15.29_{-0.30}^{+0.41}$ & $3.24_{-0.17}^{+0.17}$ & $0.43_{-0.07}^{+0.07}$ & 8275.2/8188& & & & & &\\[5pt]
& $10.0-15.0$ & $9.72_{-0.16}^{+0.16}$ & $3.05_{-0.13}^{+0.14}$ & $0.27_{-0.04}^{+0.04}$ & 8736.9/8994& & & & & &\\[5pt]
& $15.0-20.0$ & $5.58_{-0.14}^{+0.12}$ & $2.76_{-0.16}^{+0.17}$ & $0.22_{-0.04}^{+0.05}$ & 9028.4/9197& & & & & &\\[5pt]
& $20.0-25.4$ & $3.19_{-0.07}^{+0.07}$ & $2.77_{-0.17}^{+0.18}$ & $0.29_{-0.05}^{+0.05}$ & 9238.9/9550& & & & & &\\[5pt]
& $25.4-42.2$ & $1.55_{-0.03}^{+0.03}$ & $2.02_{-0.07}^{+0.08}$ & $0.16_{-0.02}^{+0.02}$ & 11805.7/11190& & & & & &\\[5pt]
& $42.2-59.0$ & $0.54_{-0.02}^{+0.02}$ & $1.57_{-0.07}^{+0.07}$ & $0.12_{-0.02}^{+0.02}$ & 12162.6/11328& & & & & &\\[5pt]
& $59.0-74.9$ & $0.27_{-0.02}^{+0.02}$ & $1.19_{-0.06}^{+0.06}$ & $0.04_{-0.01}^{+0.02}$ & 12375.9/11358 & & \multirow{2}{*}{$59.0-90.8$} & \multirow{2}{*}{$0.21_{-0.01}^{+0.02}$} & \multirow{2}{*}{$1.15_{-0.06}^{+0.04}$} & \multirow{2}{*}{$0.05_{-0.02}^{+0.01}$} & \multirow{2}{*}{12672.2/11375}\\[5pt]
& $74.9-90.8$ & $0.14_{-0.02}^{+0.03}$ & $1.14_{-0.07}^{+0.05}$ & $0.09_{-0.04}^{+0.03}$ & 12603.0/11365& & & & & &\\[5pt]
\hline \\[-1.7ex]
\multirow{10}{*}{EAST} & $0.0-2.5$ & $175.93_{-6.73}^{+6.58}$ & $1.66_{-0.03}^{+0.03}$ & $0.78_{-0.07}^{+0.07}$ & 7326.4/7254 & \multirow{10}{*}{NORTH} & $0.0-2.5$ & $120.58_{-5.63}^{+5.70}$ & $2.09_{-0.09}^{+0.09}$ & $1.2_{-0.15}^{+0.16}$ & 7026.4/7115\\[5pt]
& $2.5-5.0$ & $55.04_{-2.03}^{+2.01}$ & $2.91_{-0.20}^{+0.23}$ & $0.69_{-0.11}^{+0.14}$ & 7227.7/7289 & & $2.5-5.0$ & $44.92_{-1.85}^{+1.85}$ & $3.35_{-0.26}^{+0.30}$ & $0.82_{-0.16}^{+0.19}$ & 7114.3/7214\\[5pt]
& $5.0-7.5$ & $30.66_{-1.10}^{+1.13}$ & $3.24_{-0.26}^{+0.29}$ & $0.42_{-0.11}^{+0.12}$ & 7183.9/7238 & & $5.0-7.5$ & $27.06_{-1.0}^{+1.04}$ & $2.64_{-0.22}^{+0.25}$ & $0.37_{-0.08}^{+0.10}$ & 7157.7/7158\\[5pt]
& $7.5-10.0$ & - & - & - & - & & $7.5-10.0$ & $15.64_{-0.71}^{+1.01}$ & $3.01_{-0.33}^{+0.31}$ & $0.50_{-0.15}^{+0.15}$ & 6970.3/7127\\[5pt]
& $10.0-15.0$ & $12.73_{-0.53}^{+0.40}$ & $3.22_{-0.25}^{+0.33}$ & $0.35_{-0.08}^{+0.12}$ & 7749.6/7743 & & $10.0-15.0$ & $8.88_{-0.30}^{+0.30}$ & $3.10_{-0.25}^{+0.31}$ & $0.35_{-0.09}^{+0.10}$ & 7526.0/7630\\[5pt]
& $15.0-20.0$ & $7.57_{-0.35}^{+0.25}$ & $3.18_{-0.29}^{+0.41}$ & $0.36_{-0.09}^{+0.12}$ & 7642.4/7814 & & $15.0-20.0$ & $5.02_{-0.30}^{+0.26}$ & $2.43_{-0.24}^{+0.31}$ & $0.16_{-0.06}^{+0.09}$ & 7495.6/7586\\[5pt]
& $20.0-25.4$ & $4.62_{-0.17}^{+0.17}$ & $2.87_{-0.26}^{+0.32}$ & $0.30_{-0.08}^{+0.10}$ & 7785.2/7840 & & $20.0-25.4$ & $2.92_{-0.16}^{+0.15}$ & $2.35_{-0.26}^{+0.31}$ & $0.28_{-0.08}^{+0.11}$ & 7552.5/7736\\[5pt]
& $25.4-42.2$ & $2.03_{-0.06}^{+0.06}$ & $2.25_{-0.16}^{+0.18}$ & $0.19_{-0.04}^{+0.05}$ & 9244.7/9544 & & $25.4-42.2$ & $1.43_{-0.07}^{+0.07}$ & $2.09_{-0.16}^{+0.21}$ & $0.20_{-0.05}^{+0.07}$ & 9038.9/9508\\[5pt]
& $42.2-59.0$ & $0.74_{-0.07}^{+0.06}$ & $1.57_{-0.17}^{+0.27}$ & $0.03_{-0.03}^{+0.04}$ & 9336.8/9906 & & $42.2-59.0$ & $0.41_{-0.04}^{+0.04}$ & $1.59_{-0.11}^{+0.16}$ & $0.24_{-0.08}^{+0.11}$ & 9452.5/9995\\[5pt]
& $59.0-90.8$ & $0.12_{-0.03}^{+0.03}$ & $1.15_{-0.08}^{+0.06}$ & $0.13_{-0.06}^{+0.08}$ & 11774.3/11145 & & $59.0-90.8$ & $0.20_{-0.02}^{+0.03}$ & $1.22_{-0.11}^{+0.07}$ & $0.05_{-0.03}^{+0.03}$ & 11870.2/11164\\[5pt]
\hline \\[-1.7ex]
\multirow{10}{*}{WEST} & $0.0-2.5$ & $129.66_{-5.81}^{+5.85}$ & $2.19_{-0.10}^{+0.08}$ & $1.47_{-0.18}^{+0.17}$ & 7262.4/7234 & \multirow{10}{*}{SOUTH} & $0.0-2.5$ & $167.64_{-6.33}^{+6.23}$ & $1.69_{-0.03}^{+0.04}$ & $0.76_{-0.07}^{+0.07}$ & 7257.7/7216\\[5pt]
& $2.5-5.0$ & $38.27_{-1.73}^{+1.79}$ & $2.78_{-0.21}^{+0.22}$ & $1.11_{-0.18}^{+0.20}$ & 7104.9/7164 & & $2.5-5.0$ & $37.66_{-1.66}^{+1.90}$ & $2.99_{-0.24}^{+0.27}$ & $0.71_{-0.15}^{+0.17}$ & 7000.4/7077\\[5pt]
& $5.0-7.5$ & $21.17_{-1.15}^{+0.96}$ & $3.11_{-0.33}^{+0.41}$ & $0.47_{-0.13}^{+0.18}$ & 7000.1/7090 & & $5.0-7.5$ & $22.05_{-1.08}^{+0.79}$ & $3.33_{-0.41}^{+0.51}$ & $0.02_{-0.02}^{+0.11}$ & 6874.6/6993\\[5pt]
& $7.5-10.0$ & $13.84_{-0.88}^{+0.67}$ & $3.14_{-0.33}^{+0.44}$ & $0.55_{-0.14}^{+0.22}$ & 7028.4/7050 & & $7.5-10.0$ & $12.48_{-0.61}^{+0.85}$ & $3.57_{-0.52}^{+0.67}$ & $0.15_{-0.14}^{+0.16}$ & 6897.2/6933\\[5pt]
& $10.0-15.0$ & $9.42_{-0.32}^{+0.32}$ & $2.93_{-0.25}^{+0.29}$ & $0.25_{-0.07}^{+0.08}$ & 7605.9/7646 & & $10.0-15.0$ & $7.87_{-0.37}^{+0.42}$ & $3.03_{-0.34}^{+0.36}$ & $0.14_{-0.09}^{+0.10}$ & 7410.4/7493\\[5pt]
& $15.0-20.0$ & $5.47_{-0.22}^{+0.22}$ & $2.33_{-0.19}^{+0.22}$ & $0.2_{-0.06}^{+0.07}$ & 7504.7/7558 & & $15.0-20.0$ & $4.39_{-0.32}^{+0.24}$ & $3.41_{-0.51}^{+0.98}$ & $0.15_{-0.07}^{+0.20}$ & 7510.2/7540\\[5pt]
& $20.0-25.4$ & $2.93_{-0.21}^{+0.18}$ & $2.66_{-0.32}^{+0.49}$ & $0.28_{-0.10}^{+0.16}$ & 7575.9/7764 & & $20.0-25.4$ & $2.31_{-0.12}^{+0.13}$ & $3.36_{-0.5}^{+0.62}$ & $0.36_{-0.15}^{+0.19}$ & 7378.6/7586\\[5pt]
& $25.4-42.2$ & $1.39_{-0.08}^{+0.08}$ & $2.39_{-0.22}^{+0.28}$ & $0.26_{-0.07}^{+0.10}$ & 9170.7/9576 & & $25.4-42.2$ & $1.12_{-0.07}^{+0.05}$ & $2.24_{-0.22}^{+0.31}$ & $0.18_{-0.06}^{+0.08}$ & 8949.5/9309\\[5pt]
& $42.2-59.0$ & $0.56_{-0.03}^{+0.04}$ & $1.66_{-0.16}^{+0.16}$ & $0.19_{-0.05}^{+0.06}$ & 9537.2/10045 & & $42.2-59.0$ & $0.40_{-0.05}^{+0.05}$ & $2.28_{-0.48}^{+0.91}$ & $0.18_{-0.11}^{+0.23}$ & 9410.9/9826\\[5pt]
& $59.0-90.8$ & $0.20_{-0.02}^{+0.02}$ & $1.31_{-0.12}^{+0.10}$ & $0.11_{-0.04}^{+0.04}$ & 11961.9/11166 & & $59.0-90.8$ & $0.27_{-0.04}^{+0.05}$ & $1.08_{-0.11}^{+0.12}$ & $0.02_{-0.02}^{+0.03}$ & 11606.6/11109\\[5pt]
\hline
\hline
    \end{tabular}}
    \label{tab:spectro}
\end{table*}

\begin{table*}
    \centering
    \caption{Spectral analysis results (2T fit) of Centaurus cluster using eRASS:5. Subscript h and c are for hotter and cooler components, respectively. The metallicity, $Z$, is linked across both components.}
    \resizebox{\textwidth}{!}{\begin{tabular}{c c c c c c c c}
    \hline
    \hline
bin & $norm_\mathrm{h}$ & $k_\mathrm{B}T_\mathrm{h}$ & $norm_\mathrm{c}$ & $k_\mathrm{B}T_\mathrm{c}$ & $Z$ & \multirow{2}{*}{$norm_\mathrm{h}/norm_\mathrm{c}$} & \multirow{2}{*}{stat/dof}\\
$[\mathrm{arcmin}]$ & $[10^{-5}$ cm$^{-5}/$arcmin$^2]$ & $[$keV$]$ & $[10^{-5}$ cm$^{-5}/$arcmin$^2]$ & $[$keV$]$ & $[Z_{\odot}]$ & & \\
\hline \\[-1.7ex]
$0.0-0.8$ & $249.44_{-16.51}^{+16.87}$ & $1.89_{-0.11}^{+0.08}$ & $34.8_{-6.73}^{+7.71}$ & $1.0_{-0.06}^{+0.05}$ & $1.59_{-0.18}^{+0.22}$ & 7.2 & 7300.9/7292\\[5pt]
$0.8-1.7$ & $101.77_{-32.10}^{+22.30}$ & $3.51_{-0.39}^{+1.75}$ & $50.09_{-27.65}^{+38.44}$ & $1.56_{-0.18}^{+0.14}$ & $1.64_{-0.21}^{+0.29}$ & 2.0 & 7523.9/7496\\[5pt]
$1.7-2.5$ & $87.17_{-4.75}^{+3.85}$ & $2.39_{-0.11}^{+0.16}$ & $0.95_{-0.53}^{+1.35}$ & $0.95_{-0.20}^{+0.30}$ & $1.15_{-0.13}^{+0.19}$ & 91.9 & 7439.7/7431\\[5pt]
\hline
\hline
\end{tabular}}
\label{tab:spectro2T}
\end{table*}

\end{appendix}

\end{document}